\documentstyle[prl,aps]{revtex}
\input BoxedEPS.tex
\SetOzTeXEPSFSpecial
\HideDisplacementBoxes

\begin{document}

\twocolumn[\hsize\textwidth\columnwidth\hsize
           \csname @twocolumnfalse\endcsname
\title{Magnetic evidence for hot superconductivity in 
multi-walled carbon nanotubes} 
\author{Guo-meng Zhao$^{*}$ and Pieder 
Beeli} \address{Department of Physics and Astronomy, California State 
University, Los Angeles, CA 90032, USA}

\maketitle
\widetext

\maketitle
\widetext

\begin{abstract}

We report magnetic measurements up to 1200 K on three different 
multi-walled carbon nanotube mat samples using Quantum Design 
vibrating sample magnetometers.  
Three different samples prepared from arc discharge or chemical vapor 
deposition contain magnetic impurities ranging from about 100 ppm to 
about 1.5$\%$.  Our precise magnetic data clearly 
show two superconducting transitions, one at temperatures between  
533 K and 700 K, and another at about 1200 K.  The first transition 
temperature $T_{cJ}$, which coincides with the transition temperature 
seen in the resistance data, depends very strongly on the magnetic 
field, as expected from the onset of intergrain Josephson coupling in 
granular superconductors.  The strong field dependence of $T_{cJ}$ 
also 
excludes magnetic contaminants as the origin of the first 
transition.  
We also present direct and inferred diamagnetic Meissner fractions of 
2 and 14$\%$, respectively.
The present results provide compelling evidence for superconductivity 
well above room 
temperature in multi-walled carbon nanotubes.
~\\~\\
\end{abstract}
\narrowtext
]

\section{Introduction}
It is generally believed that the superconducting 
transition temperature $T_{c}$ cannot be higher than 30 K within the 
conventional phonon-mediated mechanism. Alexandrov and Mott have 
demonstrated that the Bose-Einstein condensation of bipolarons can 
explain high-temperature superconductivity in cuprates \cite{Alex}.  
Ginzburg \cite{Ginzburg} and Little \cite{Little} have proposed that 
high-temperature superconductivity could be realized by exchanging 
high-energy electronic excitations such as excitons and plasmons.  Lee 
and Mendoza have shown that superconductivity as high as 500~K can be 
reached through a pairing interaction mediated by undamped acoustic 
plasmon modes in a quasi-one-dimensional (1D) electronic system 
\cite{Lee}.  Moreover, high-temperature superconductivity can occur in 
a multi-layer electronic system due to an attraction of charge 
carriers in the same conducting layer via exchange of virtual plasmons 
in neighboring layers \cite{Cui}.  If these plasmon-mediated pairing 
mechanisms are relevant, one should be able to find high-temperature 
superconductivity in quasi-one-dimensional and/or multi-layer systems 
such as cuprates, carbon nanotubes (CNTs), and graphites.

Carbon nanotubes constitute a novel class of quasi-one-dimensional 
materials which offer the potential for high-temperature 
superconductivity.  The simplest single-walled nanotube (SWNT) 
consists of a single graphite sheet which is curved into a long 
cylinder, with a diameter which can be smaller than 1 nm.  
Band-structure calculations predict that carbon nanotubes have two 
types of electronic structures depending on the chirality 
\cite{Saito1,Ajiki}, which is indexed by a chiral vector $(n,m)$: $n 
-m = 3I +\nu$, where $I$, $n$, $m$ are the integers, and $\nu$ = 0, 
$\pm 1$.  The tubes with $\nu$ = 0 are metallic while the tubes with 
$\nu$ = $\pm$1 are semiconducting.  For metallic chirality SWNTs, there 
are two and six transverse conduction channels when the Fermi level is 
crossing the first and second subbands, respectively.  Multiwalled 
nanotubes (MWNTs) consist of at least two concentric shells which can 
have different chiralities.  The outer diameters of our arc discharge 
prepared MWNTs are centered at around 10-15 nm.  MWNTs possess both 
quasi-one-dimensional and multi-layer electronic structures.  This 
unique quasi-one-dimensional electronic structure in both SWNTs and 
MWNTs make them ideal for plasmon-mediated high-temperature 
superconductivity.

In order to confirm the existence of superconductivity, it is 
essential to provide two important 
signatures: the Meissner effect and the resistive transition. In 2001 
Zhao {\em et al.} \cite{Zhao1} provided 
magnetic and 
electrical evidence for possible superconductivity above room 
temperature in MWNT mat samples.  The resistivity data show a 
possible 
superconductive transition at about 700 K. Since then, Zhao 
\cite{Zhaorev2,Zhaorev4} has analyzed previously published magnetic, 
electrical, and optical data for both single-walled and multi-walled 
carbon nanotubes, and provided over twenty arguments for the 
existence of hot superconductivity.  In 2003, Kopelevich and 
coworkers \cite{Kop} gave magnetic 
evidence for local superconductivity up to 270 K in graphite-sulfur 
composites.  The observations of hot superconductivity in carbon 
nanotubes and graphite-sulfur composites suggest a common 
microscopic origin of superconductivity.

In this article, we report magnetic measurements up to 1200 K on 
three 
different multi-walled carbon nanotube mat samples using sensitive 
vibrating 
sample magnetometers (VSMs) from Quantum Design.  The ultra low-field 
option of the VSM system allows us to do the magnetization 
measurements in a field as low as 0.03 Oe.  One sample is prepared 
from arc discharge (denoted as AD) and contains about 100 
ppm magnetic impurities.  The second sample is prepared from chemical 
vapor deposition (denoted as CVD1) and contains about 
1.5$\%$ Fe$_{3}$O$_{4}$.  The third sample is also prepared
from chemical vapor deposition (denoted as CVD2) and has about 0.3$\%$ impurities.  Our extensive magnetic data
consistently show two superconducting transitions, one at about 700 
K, 533 K, and 700 K for sample AD, sample CVD1 and sample CVD2, 
respectively, and another at about 1200 K.  The first transition 
temperature $T_{cJ}$, which coincides with the transition temperature 
seen in the previous resistance data \cite{Zhao1}, depends very strongly on the magnetic 
field in the low field range, as expected from the onset of 
intergrain 
Josephson coupling in granular superconductors.  The strong field 
dependence of $T_{cJ}$ also excludes magnetic contaminants as the 
origin of the first transition.  We also present direct and inferred 
diamagnetic Meissner fractions of 2 and 14$\%$, respectively.  Such a
large field-cooled diamagnetic susceptibility is {\em only} consistent with 
superconductivity. The superconducting remanent magnetization, which 
can be 
distinguished from the remanent contribution of magnetic impurities, exists up 
to 700 K, indicating a zero-resistance state well above room-temperature in 
some parts of the mat samples.  The present results provide compelling 
evidence for superconductivity well above room temperature in 
multi-walled carbon nanotubes.

\section{Experiment}
\subsection{Sample characterization}

Three different multi-walled carbon nanotube mat samples are obtained 
from 
SES Research of Houston.  Sample AD is prepared from an arc 
discharge process with no metal catalysts.  The multi-walled 
nanotubes consist of 5-20 graphite layers with diameters between 2 and 
20 
nm and lengths between 100 nm and 2~$\mu$m.  The tubes are naturally 
assembled into bundles, and the bundles into mats where bundles are 
entangled with each other.  Sample AD comes from the same lot as 
those studied in Ref.~\cite{Zhao1}.  This sample contains a total of 
about 100 ppm magnetic impurities, as determined from the magnetic 
measurement (see below).  The mass of sample AD is 41 mg for most 
magnetic measurements.  Sample CVD1 is prepared from chemical vapor 
deposition and contains about 1.5$\%$
Fe$_{3}$O$_{4}$.  The mass of sample CVD1 is 5 mg.  Sample CVD2 is 
also 
prepared from chemical vapor deposition and  contains a 
total of about 0.3$\%$ impurities.  The mass of sample CVD2 
is 11 mg.

\subsection{Measurement}

Magnetization was measured by two Quantum Design vibrating sample 
magnetometers (VSMs).  Extensive magnetic measurements by both us and 
the staff of Quantum Design (up to 1100 K) on a control sample of 
Er$_{2}$O$_{3}$ show that the absolute accuracy of the moment is 
better than 1.0 $\times$ 10$^{-7}$ emu at low fields after correcting 
for a constant offset of 5.0 $\times$ 10$^{-7}$ emu due to the 
interaction 
between the VSM drive head and the pick-up coils (see below).  Fields 
between 
0.03 Oe and 15 Oe are obtained by using the ultra low-field option.  
The magnitude and direction of the field profile were measured by a 
flux-gate device.  The oven option is rated for temperatures up to 
1100 K, but we successfully pushed the temperature up to 1200 K in 
three runs.  In the fourth run, the heater broke at about 1132 K.  
The sample chamber is in a high vacuum ($<$10$^{-4}$ torr) during 
measurements, which prevents the carbon nanotubes from burning.  The 
VSM oscillation frequency is 40 Hz and the oscillation amplitude is 
1~mm.  When 
the sample is inserted into the sample chamber, it first 
experiences a positive (upward) field of about 200 Oe and then the 
same negative (downward) field due to the presence of the linear 
motor 
used to vibrate the sample.  The magnetic hysteresis loop was 
measured 
for the sample holder alone (including the heater stick, Zircar 
cement, and copper foil) at 300 K.  The saturation moment is 
5.4$\times$10$^{-6}$ emu and the linear moment per Oe is 
$-$4.3$\times$10$^{-9}$ emu.  Such a saturation moment corresponds to 
about 5.4$\times$10$^{-5}$ mg of magnetic impurities if we take a 
typical saturation magnetization of 100 emu/g.  Our sample 
masses range from 5 mg to 41 mg, which are 10$^{5}$-10$^{6}$ times 
larger than the mass of magnetic impurities in the sample holder.  
Therefore, the magnetic signal from the sample holder 
is negligible compared with our sample signal.

\section{Results and discussion}

\subsection{Experimental results for a control sample of 
Er$_{2}$O$_{3}$: Determinations of measurement accuracies }

Because of the high temperature, low field and high sensitivity 
requirements of 
this work, we worked closely with the staff of Quantum Design to 
define the 
measurement accuracies in moment and temperature. Prior to this work, 
Quantum Design's VSM was rated for temperature up to 1000 K and for 
moment $<$ 7$\times$10$^{-7}$ emu. After we 
made extensive magnetic measurements on a control sample of 
Er$_{2}$O$_{3}$ up to 1100 K, 
we find that the temperature is accurate up to at least 1100 K and the
absolute accuracy in moment can be better than 1$\times$10$^{-7}$ emu 
in low fields.  We believe that the temperature reading is also 
reliable above 1100 K except that there is a risk in breaking the 
heater stick for temperatures higher than 1100 K.

Fig.~1 shows the temperature dependence of the field-cooled 
susceptibility for a 9.4 mg Er$_{2}$O$_{3}$ sample in a field of 10 
kOe.  The cooling rate for this measurement is 10 K per minute.  We 
can fit the data  by the Curie-Weiss law
\begin{equation}
\chi = \frac{C}{T + \theta},
\end{equation}
where $C$ is the Curie constant.  The solid line is the fitted 
curve which lies almost perfectly on the top of data points. The best 
fit 
gives $C$ = (6.327$\pm$0.002)$\times$10$^{-2}$ emu/g and 
$\theta$ = 24.2$\pm$0.2 K.  We also fit warm-up data by Eq.~1.  The 
best fit gives $C$ = (6.304$\pm$0.002)$\times$10$^{-2}$ emu/g and 
$\theta$ = 16.7$\pm$0.2 K.  The Curie constants deduced from the 
cool-down and warm-up data differ only by 0.36$\%$.  The $\theta$
values differ by 7.6 K,
suggesting that there is a small thermal lag of about 3.8 K for the 
heating/cooling rate of 10 K per minute.  For the 
heating/cooling rate of 100 K per minute, there is a thermal lag of 
about 20 K.  For most measurements on the nanotube samples, the 
heating/cooling rate is 50 K per minute and thus the thermal lag should be 
less than 20 K.

\begin{figure}[htb]
	\vspace{0.6cm} 
 \ForceWidth{7cm}
\centerline{\BoxedEPSF{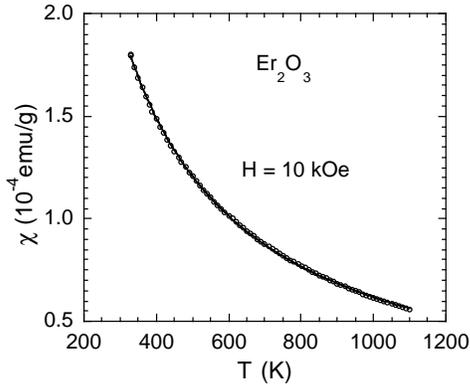}}
	\vspace{0.3cm}
\caption[~]{The temperature dependence of the field-cooled 
susceptibility for a 9.4 mg Er$_{2}$O$_{3}$ sample in a 
field of 10 kOe. The data shown are thinned for clarity. The solid line is 
a fit to all the data points by the 
Curie-Weiss law.  }
\end{figure}

The intrinsic $\theta$ value for the Er$_{2}$O$_{3}$ sample is 
determined to be 20.5 K by taking a simple average of the two $\theta$ 
values.  From the Curie constants, we can easily calculate the 
effective magneton number of Er$^{3+}$ to be 9.8, which is about 
2$\%$ larger
than the theoretical prediction (9.6). This small discrepancy may 
arise from the uncertainty in the sample weight.

Since the moments of the Er$_{2}$O$_{3}$ sample in 10 kOe are in the 
range 
between 0.005-0.016 emu, 
which are far above the instrument resolution, the data must be very 
accurate. Therefore, by comparing these accurate data with those data 
taken in 
the low field range, we can determine the absolute accuracy of the 
moment. In Fig.~2, we show the temperature dependencies of the offset 
moments 
in the fields of 0.06 Oe and 4.21 Oe, respectively. The data shown are thinned for clarity. The offset 
moments 
are obtained by subtracting the above determined Curie-Weiss term 
from 
the raw data.  In the field of 0.06 Oe, the Curie-Weiss term is very 
small 
(e.g., 1.17$\times$10$^{-7}$ emu at 300 K), so the offset moments  
dominate in the whole temperature region.  Since the data are taken 
with 3 second per point averaging, there is significant data 
scattering.  We can average the data by using an 8th order 
polynominal 
fit to all the data points (dotted line).  The averaged curve shows no sizable temperature 
dependence.  Then we simply fit the data by a constant line (solid line).  The 
best 
fits to the 0.06 Oe and 4.21 Oe data give $m_{offset}$ = 
(4.54$\pm$0.07)$\times$10$^{-7}$ emu and 
(5.20$\pm$0.05)$\times$10$^{-7}$ emu, respectively.

\begin{figure}[htb] 
 \ForceWidth{7cm}
\centerline{\BoxedEPSF{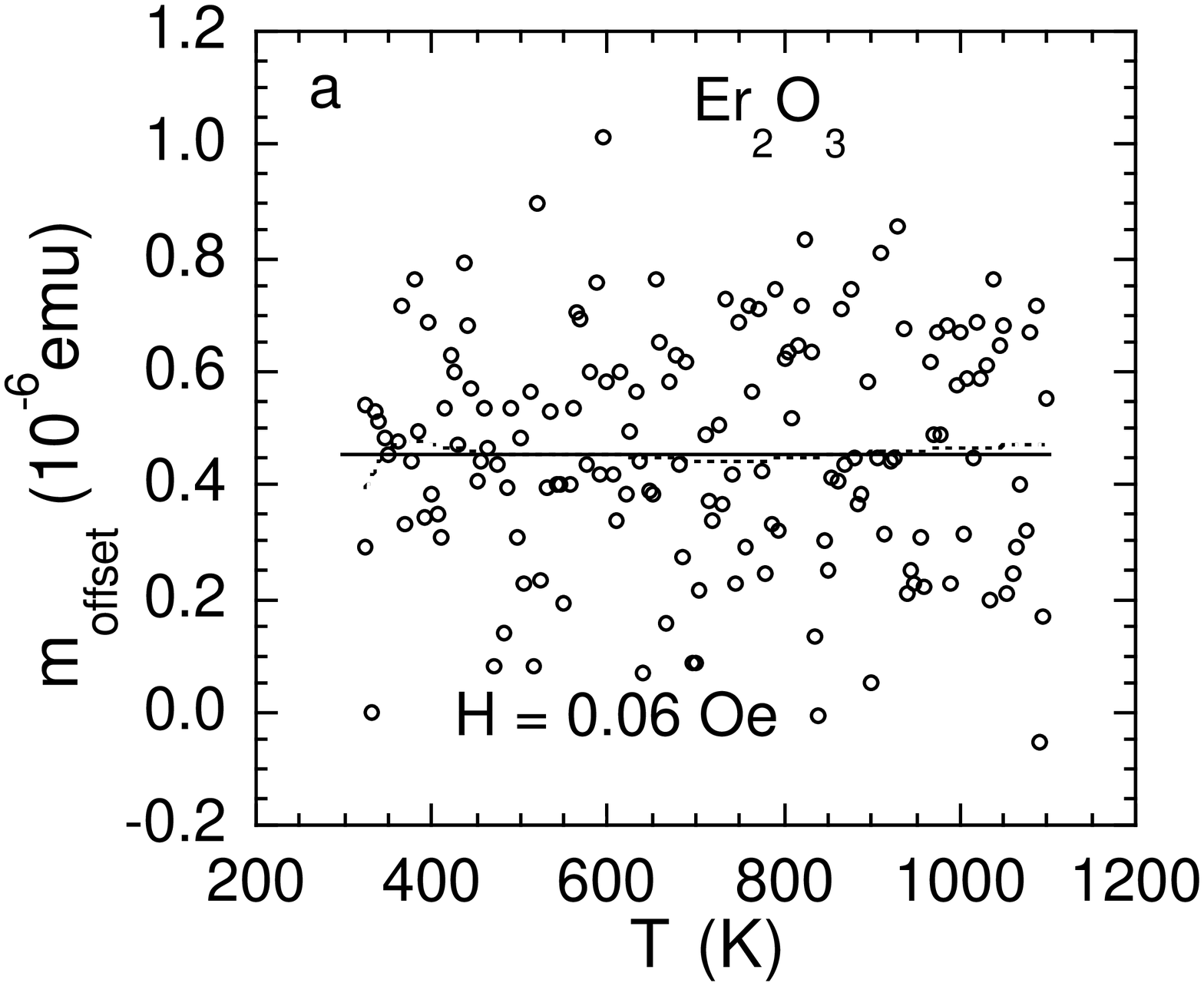}}
\vspace{-0.1cm}
\ForceWidth{7cm}
\centerline{\BoxedEPSF{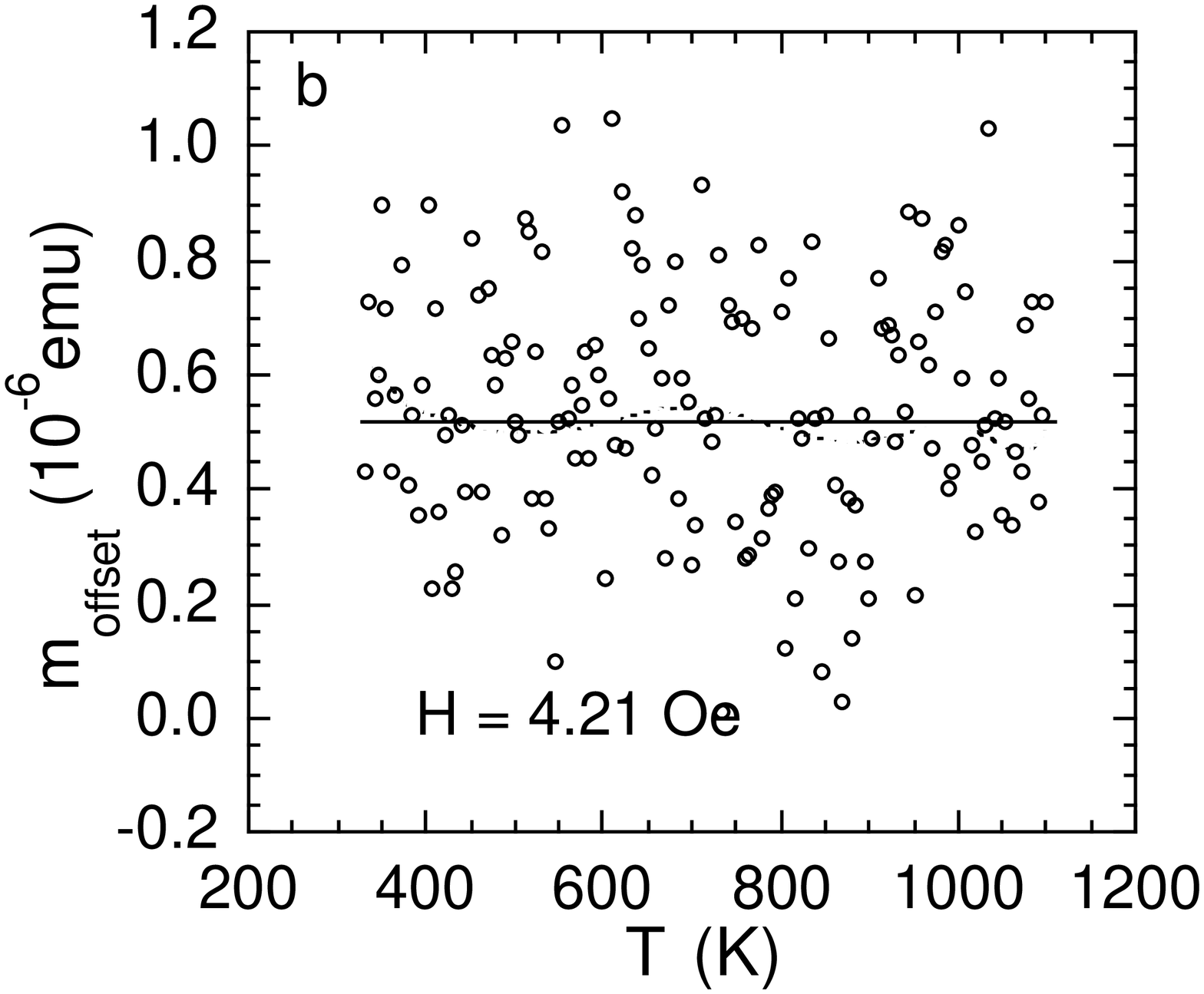}}
	\vspace{0.3cm}
\caption[~]{The temperature dependencies of the offset 
moments in the fields of a) 0.06 Oe and b) 4.21 Oe, respectively. The data 
shown are thinned for clarity. The offset 
moments 
are obtained by subtracting the above determined Curie-Weiss term 
from the raw data. }
\end{figure}

Fig.~3 shows the field dependence of the offset moment. The offset 
moments determined 
from both cool-down and warm-up data are included in the figure 
for comparison. It is clear that the offset moments are centered 
around 5.0$\times$10$^{-7}$ emu in this low-field range (0.06-4.21 
Oe).  The maximum deviation from 5.0$\times$10$^{-7}$ emu is 
0.46$\times$10$^{-7}$ emu.  This offset moment has negligible field 
and temperature dependencies relative to our signals of interest and is 
caused by the interaction between the VSM drive head and the pick-up 
coils.  Therefore, if we correct low-field moments by the constant 
offset moment, the absolute accuracy of the corrected moments should 
be better than 0.6$\times$10$^{-7}$ emu.  In other words, if we use a 
proper averaging for data points and correct for the constant offset 
of 5.0$\times$10$^{-7}$ emu, the absolute accuracy of the corrected 
moment can be better than 1$\times$10$^{-7}$ emu.  We thus correct all 
the following data by the constant offset of 5.0$\times$10$^{-7}$ emu.  
Since the following data were collected with 10 second per point 
averaging, the data scattering is significantly reduced.

\begin{figure}[htb] 
 \ForceWidth{7cm}
\centerline{\BoxedEPSF{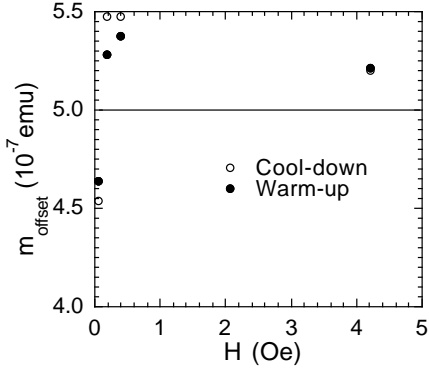}}
	\vspace{0.3cm}
\caption[~]{The field dependence of the offset moment determined by 
fitting both cool-down and warm-up data. the offset moments are centered 
around 5.0$\times$10$^{-7}$ emu. 
The maximum deviation from 5.0$\times$10$^{-7}$ emu is 
0.46$\times$10$^{-7}$ emu.  The constant offset moment is caused by the interaction 
between the VSM drive head and the pick-up coils.  }
\end{figure}

\subsection{Superconductivity in AD prepared samples}

The concentration of magnetic contaminants in AD prepared samples 
should be very small since the samples were prepared by arc discharge 
process without metal catalysts.  Nevertheless, a very small amount of 
magnetic impurities could have a dominant contribution to magnetic 
properties of a sample if the intrinsic magnetic susceptibility of the 
sample is small.  Therefore, it is important to accurately determine 
the concentration of magnetic contaminants in the AD prepared samples 
before we can draw any meaningful conclusions from the magnetic data.

Fig.~4a shows the field-cooled 
susceptibility in 100 Oe for a 14.6 mg AD prepared sample.  The sample 
was heated up to 1000 K in 100 Oe and cooled down from 1000 K in the 
same field.  At about 875 K, the susceptibility increases sharply, 
indicating ferrimagnetic/ferromagnetic ordering with a Curie 
temperature ($T_{C}$) of about 875 K.  From the Curie temperature, we 
can conclude that this AD prepared sample is contaminated with Fe$_{3}$O$_{4}$ magnetic impurities.  At 1000 K, we also see that the 
susceptibility increases rapidly with time.  This could be caused by 
a thermal lag and/or by the reduction of Fe$_{3}$O$_{4}$ to Fe which has a Curie 
temperature of 1043 K.  The ferrimagnetic component can 
be estimated by subtracting the extrapolated line in the figure from 
the total susceptibility.  In Fig.~4b we plot the estimated 
ferrimagnetic component of the susceptibility.  The temperature 
dependence of the susceptibility appears to be in good agreement with 
that for a typical ferromagnet.  At 600 K, the ferrimagnetic 
susceptibility of the Fe$_{3}$O$_{4}$ impurities is about 
3.0$\times$10$^{-6}$ emu per gram of MWNT sample.

We can fit the ferrimagnetic component of the susceptibility  near 
$T_{C}$ by an equation:
\begin{equation}
\chi_{FM} (T) = \chi_{FM} (0)[1- (T/T_{C})^{p}].
\end{equation}
The best fit gives $T_{C}$ = 876.7~K, $p$ = 23.75, and $\chi_{FM} (0)$ = 
2.5$\times$10$^{-6}$ emu/g.  It is clear that the fitted curve is 
significantly underestimated at temperatures well below $T_{C}$.

\begin{figure}[htb] 
 \ForceWidth{7cm}
\centerline{\BoxedEPSF{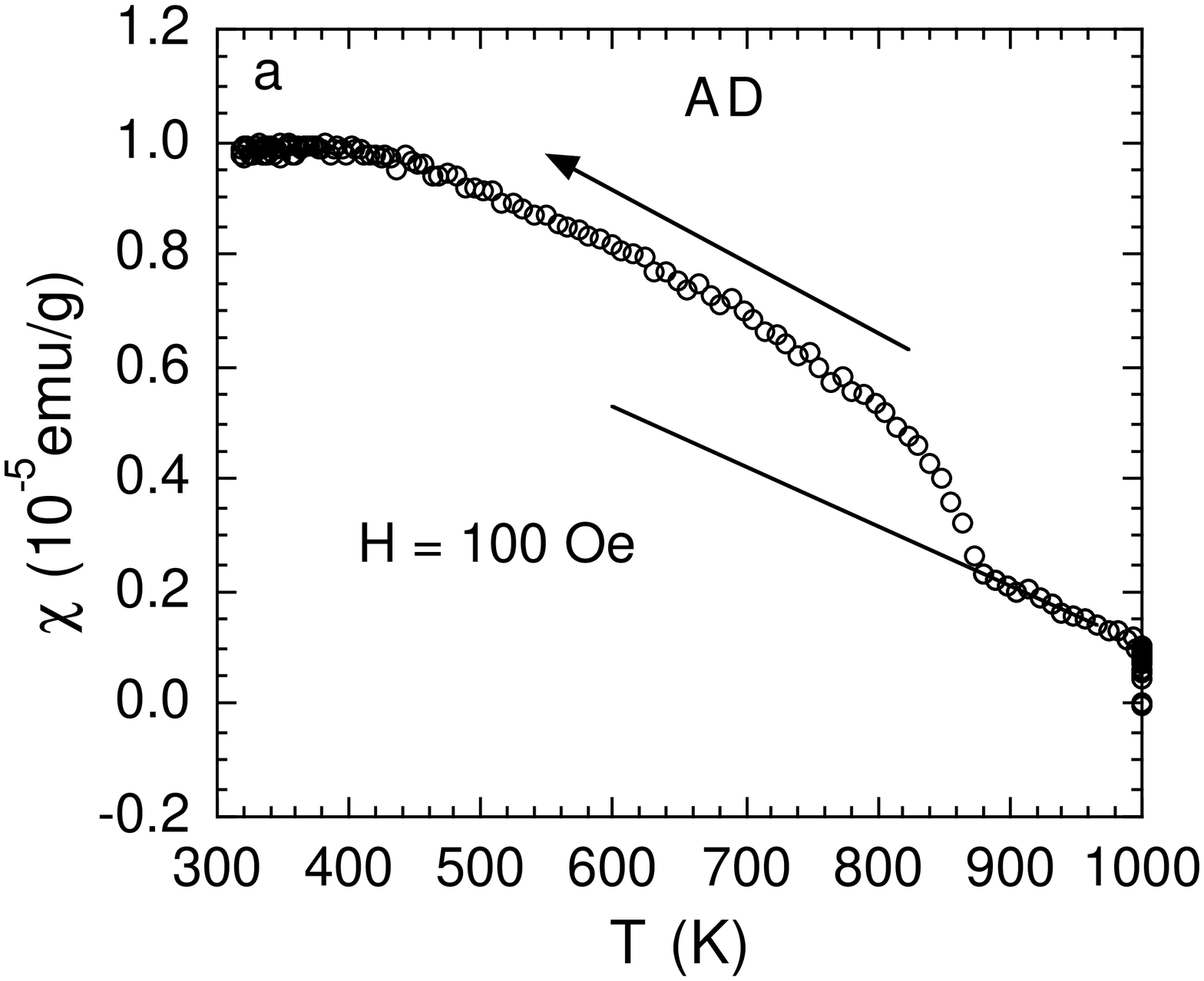}}
\vspace{-0.1cm}
\ForceWidth{7cm}
\centerline{\BoxedEPSF{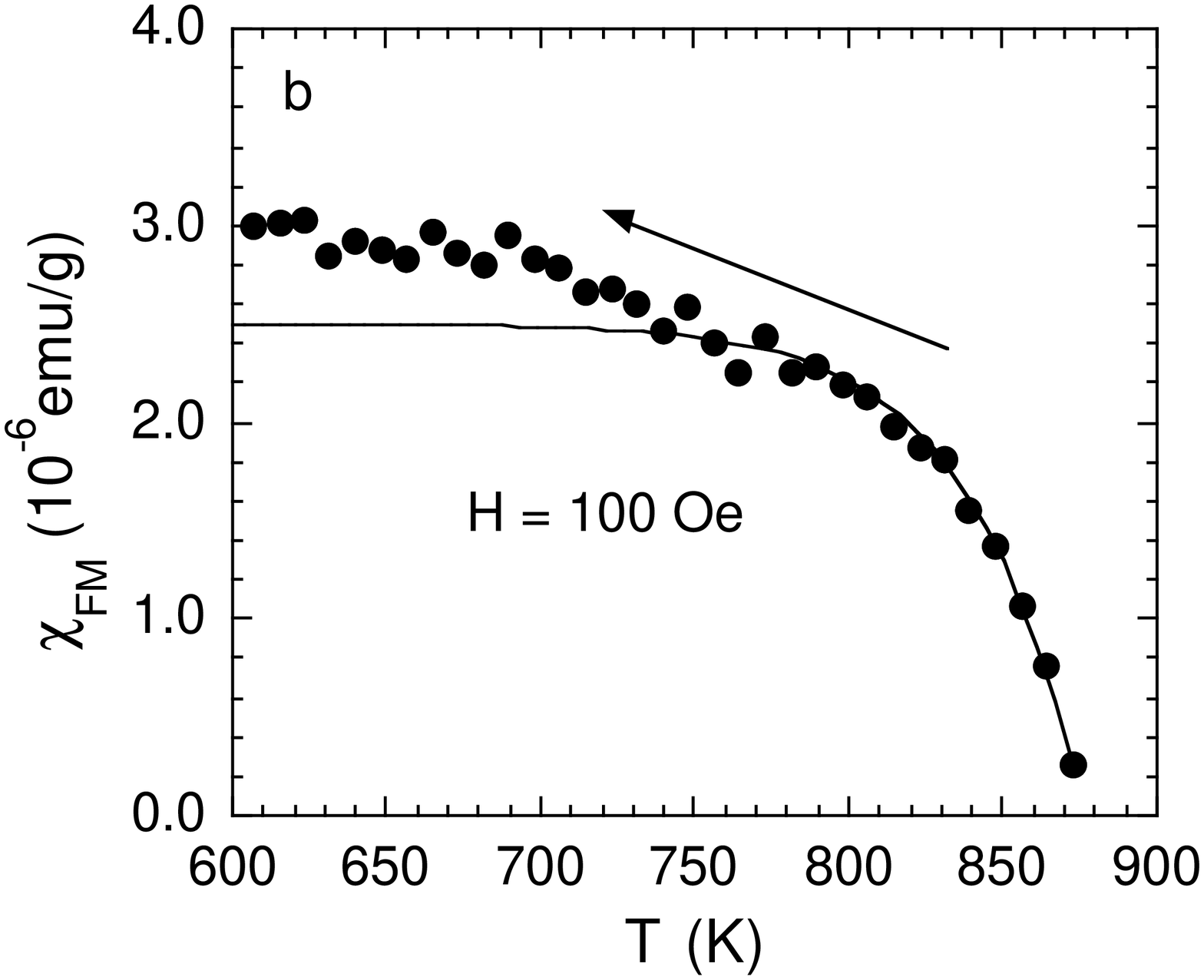}}
	\vspace{0.3cm}
\caption[~]{a) The field-cooled susceptibility in 100 Oe for a 14.6 mg 
AD prepared sample.  b) The estimated ferrimagnetic component of the 
susceptibility due to magnetic impurities.  The data are fitted by 
Eq.~2 with $T_{C}$ = 876.7~K, $p$ = 23.75, and $\chi_{FM} (0)$ = 
2.5$\times$10$^{-6}$ emu/g. }
\end{figure}

We remind ourselves that for materials 
lightly contaminated with magnetic impurities having large 
permeabilities, the low field susceptibility is independent of the 
permeability and depends only on the shape (demagnetization factor) 
and concentration of the contaminant.  The shape effect is averaged 
out through a random orientation so that the average demagnetization 
factor $N$ should be about 1/3 and the low-field susceptibility 
should be $1/(4N\pi$) = 0.24 emu/cm$^{3}$.  Since the specific weight 
of 
Fe$_{3}$O$_{4}$ is 5.3 g/cm$^{3}$, we 
easily calculate the low-field susceptibility of pure Fe$_{3}$O$_{4}$ 
particles to be 0.045 emu/g.  The specific weight of Fe is 7.9 
g/cm$^{3}$, so the low-field susceptibility of pure Fe particles is 
0.030 emu/g.  This implies that the concentration of the 
Fe$_{3}$O$_{4}$ impurities in the AD prepared sample is about 
3$\times$10$^{-6}$/0.045 = 67 ppm. If the step-like increase of the 
susceptibility at 1000 K (1.2$\times$10$^{-6}$ emu/g) could arise from 
the reduction of Fe$_{3}$O$_{4}$ to Fe, the concentration of the 
reduced Fe$_{3}$O$_{4}$ would be about 27 ppm.  Then the upper limit of 
the total Fe$_{3}$O$_{4}$ impurity concentration is about 94 ppm.  
If the Fe$_{3}$O$_{4}$ impurities are completely reduced to Fe, the 
upper limit of the Fe impurities is about 66 ppm.

\begin{figure}[htb] 
 \ForceWidth{7cm}
\centerline{\BoxedEPSF{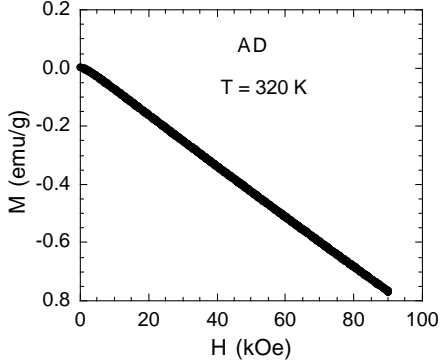}}
	\vspace{0.3cm}
\caption[~]{The field dependence of the magnetization at 320 K for the 14.6 mg
AD prepared sample. Between 20 kOe and 
90 kOe, the data can be excellently fitted by a linear relation $M = 
M_{s} +\chi_{dia}H$ with $M_{s}$ = 9.27$\times$10$^{-3}$ emu/g and 
$\chi_{dia}$ = $-$8.66$\times$10$^{-6}$ emu/g.}
\end{figure}

The total concentration of ferromagnetic and/or 
ferrimagnetic impurities in the AD prepared sample can also be 
estimated 
from the field dependence of the magnetization at 320 K, as shown in 
Fig.~5.  Above 20 kOe, the magnetization of any ferromagnetic 
impurity 
should be saturated and independent of the field.  Between 20 kOe and 
90 kOe, the data can be excellently fitted by a linear relation $M = 
M_{s} +\chi_{dia}H$ with $M_{s}$ = 9.27$\times$10$^{-3}$ emu/g and 
$\chi_{dia}$ = $-$8.66$\times$10$^{-6}$ emu/g.  The field independent 
diamagnetic susceptibility in the high-field region could be 
consistent with granular superconductivity, as observed in the 
granular 
superconductor Tl$_{2}$Ba$_{2}$CuO$_{6+y}$ 
with $T_{c}$ = 15 K (see Fig.~7 of Ref.~\cite{Berg}).  If individual 
multi-walled carbon nanotubes are superconductors, it is natural that 
multi-walled nanotube mat samples are granular superconductors.  
Given that the saturation magnetization of a typical 
ferromagnet/ferrimagnet is about 100 emu/g, the deduced saturation 
magnetization from the fit corresponds to about 100 ppm magnetic 
impurities.

Fig.~6 shows the temperature dependence of the magnetization in a 
field of 0.03 Oe for an identical AD prepared sample, but with a mass 
of 41 mg.  When the sample is inserted into the sample chamber with a 
field of 0.03 Oe, it first experiences a positive (upward) field of 
about 200 Oe and then the same negative (downward) field due to the 
presence of the linear motor used to vibrate the sample.  Therefore the 
warm-up magnetization is the sum of the negative remanent 
magnetization and the non-remanent magnetization in 0.03 Oe.  It is 
clear that the negative magnetization flattens out above 700 K and 
that there is a nearly temperature-independent negative magnetization 
between 700 K and 900 K.  Above 900 K, the magnetization increases 
sharply, reaches a peak at about 1142 K and then drops sharply between 
1142 K and 1170 K.

\begin{figure}[htb] 
 \ForceWidth{7cm}
\centerline{\BoxedEPSF{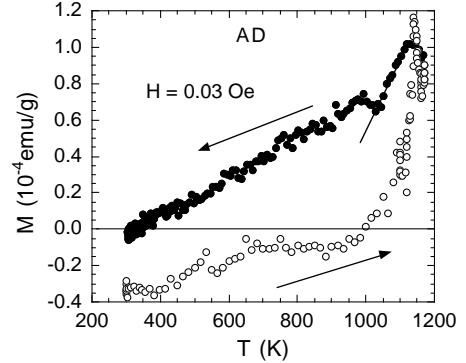}}
	\vspace{0.3cm}
\caption[~]{a) The temperature dependence of the magnetization in a field 
of 0.03 Oe for a 41 mg AD prepared MWNT mat sample.}
\end{figure}

When the sample is cooled from 1170 K, the 
magnetization does not go back to the warm-up curve, showing a 
significant hysteresis. At about 1030 K, we see a dip in the 
cool-down magnetization, which could reveal the onset of the 
ferromagnetic ordering of Fe.  The Curie temperature of Fe determined 
from these FC data is about 10 K lower than the expected value (1043 
K).  This is caused by a thermal lag of about 10 K for the cooling 
rate of 50 K per minute.  At 875 K, we do not see an anomaly 
associated with ferrimagnetic ordering of the Fe$_{3}$O$_{4}$ 
impurities.  It is very likely that the Fe$_{3}$O$_{4}$ impurities 
have almost completely been reduced into Fe after the sample was 
heated up to 1170 K at a high vacuum.

The ferromagnetic component (due to the Fe impurities) in the cool-down data 
can be estimated by subtracting the extrapolated line in Fig.~6 from 
the total magnetization.  In Fig.~7a we plot the estimated 
ferromagnetic component of the susceptibility.  The solid line is the 
fitted curve by Eq.~2.  The best fit gives $T_{C}$ = 1030.6 K, $p$ = 
14.56, and $\chi_{FM}(0)$ = 1.46$\times$10$^{-3}$ emu/g.  In Fig.~7b, we 
show the intrinsic FC susceptibility of the carbon nanotubes after 
subtracting the fitted curve of Fig.~7a from the total susceptibility.  
We see that the susceptibility drops by about 5.0$\times$10$^{-3}$ 
emu/g when the temperature is lowered from 1130 K to 305 K.  This 
implies that the magnitude of the diamagnetic component at 305 K is at 
least 5.0$\times$10$^{-3}$ emu/g, which corresponds to about 14$\%$ of 
the
full Meissner effect ($-$1/4$\pi$) given that the specific weight of 
multi-walled nanotubes is 2.17 g/cm$^{3}$.  It is interesting to note 
that the diamagnetic susceptibility at 305 K is about 
$-$1.55$\times$10$^{-3}$ emu/g, corresponding to about 4$\%$
of the full Meissner effect.  This large FC susceptibility is {\em 
only} consistent 
with superconductivity.

This temperature dependence of the intrinsic FC susceptibility is 
similar to that observed in a Nb disk (see Fig.~1 of \cite{Pust}).  
Such a temperature dependence can arise from the interplay between 
the 
diamagnetic screening currents (diamagnetic Meissner effect) and the 
paramagnetic circulating currents (paramagnetic Meissner effect) 
around the vortices \cite{MQB}.  The diamagnetic component~of
\begin{figure}[htb] 
 \ForceWidth{7cm}
\centerline{\BoxedEPSF{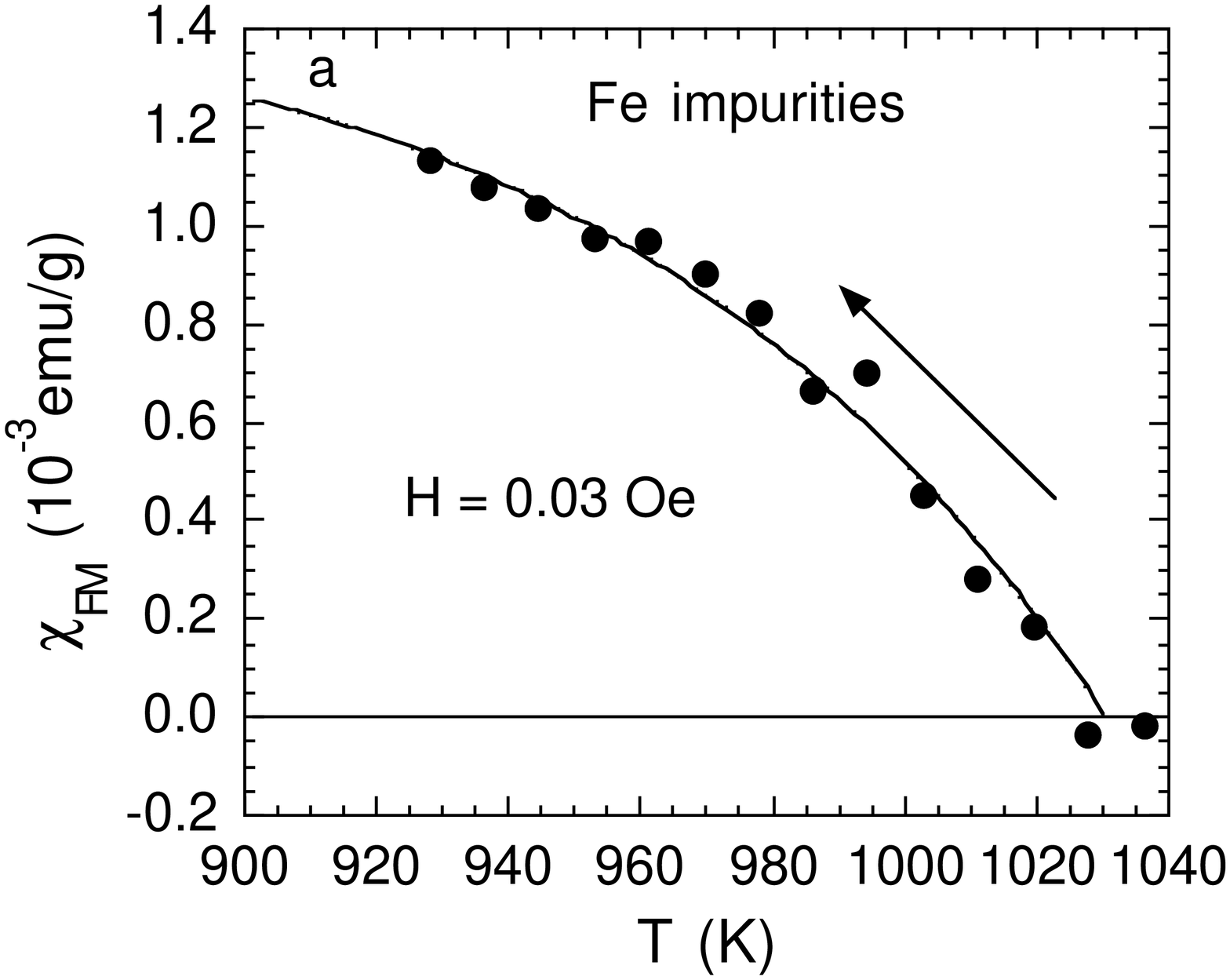}}
\vspace{-0.1cm}
\ForceWidth{7cm}
\centerline{\BoxedEPSF{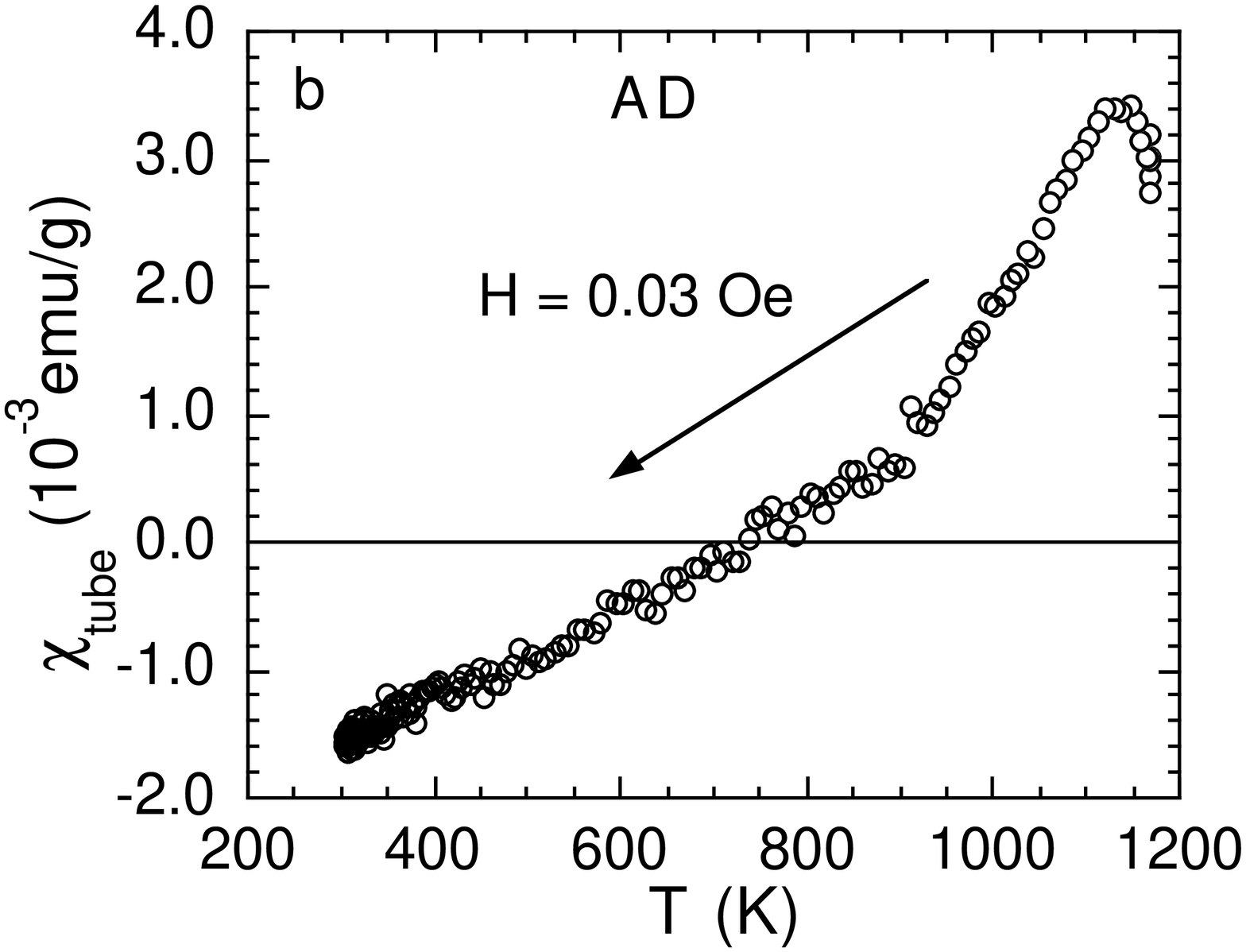}}
	\vspace{0.3cm}
\caption[~]{a) The estimated FC ferromagnetic susceptibility due to Fe 
impurities in the AD prepared MWNT mat 
sample.  b) The intrinsic FC susceptibility of the carbon nanotubes 
after subtracting the fitted curve of Fig.~7a from the total 
susceptibility. The 
diamagnetic susceptibility at 305 K is about $-$1.55$\times$10$^{-3}$ 
emu/g, corresponding to about 4$\%$
of the full Meissner effect ($-1/4\pi$). }
\end{figure}
\noindent
the 
susceptibility should be nearly independent of the magnetic field $H$ 
when $H$ is below the lower critical field $H_{c1}$ of a bulk 
sample.  
This is indeed the case for the Nb disk \cite{Pust}.  On the other 
hand, for a granular superconductor, the intergrain lower critical 
field 
tends to zero due to a very large intergrain penetration depth.  In 
the low field region, the magnitude of the diamagnetic susceptibility 
is found to increase with decreasing field much faster than $1/H$ 
(see 
Ref.~\cite{Berg,Wen,Klamut}) so that the diamagnetic component can be 
significant only in the low field range.

Moshchalkov, Qiu, and Bruyndoncx (MQB) \cite{MQB} suggested that the 
paramagnetic 
Meissner effect (PME) can be caused 
by the persistence of the giant vortex state with the fixed orbital 
quantum number $L>$ 0.  
This state is formed in any finite-size 
superconductor at the third surface critical field $H_{c3}$, which is 
higher than 
the bulk upper critical field $H_{c2}$ for smooth and clean 
surfaces.  
According to this model \cite{MQB}, the PME in small diameter 
superconducting wires is 
particularly strong because of the larger 
surface to volume ratio.  This model can naturally explain why a Pb 
nanowire 
array with an average diameter of 40 nm is paramagnetic down to 2 K 
while another 
Pb nanowire array with a diameter of 60 nm is diamagnetic below about 
3.8 K in the FC condition~\cite{Yuan}.

From the fitted curve in Fig.~7a, we find that the ferromagnetic 
susceptibility at low temperatures is about 1.46$\times$10$^{-3}$ 
emu/g.  If there were no superconductivity in the nanotubes, this 
ferromagnetic susceptibility would be too large to be consistent with 
the above determined upper limit of the Fe impurity concentration (66 ppm).  Without superconductivity in the nanotubes, the 
predicted upper limit of the low-field ferromagnetic susceptibility would be 
0.000066$\times$0.03 emu/g = 2.0$\times$10$^{-6}$ emu/g, which is a 
factor of about 700 smaller than the deduced 1.46$\times$10$^{-3}$ emu/g.  In 
order to understand this large enhancement in the low-field 
ferromagnetic susceptibility of the Fe impurities, one must assume 
that the effective fields on the magnetic impurity sites are greatly 
enhanced compared with the applied field, that is, $H_{eff}/H$ $>$$>$ 
1.  This is possible only if the host material (carbon nanotubes) 
exhibits superconductivity above the Curie temperature of magnetic 
impurities and some fraction of the impurities are flux pinning centers.  
It is known that the effective fields on those pinning centers are 
close to the lower critical field $H_{c1}$.  If the lower critical 
field inside grains (bundles) is 100 Oe and 20$\%$
of
the magnetic impurities are pinning centers, then the 
ferromagnetic susceptibility in the field of 0.03 Oe will increase by 
a factor of 660.

The PME is mostly seen in FC data \cite{Bra}. However, 
the PME is also seen in zero-field-cooled (ZFC) data of 
granular superconductors \cite{Wen,Klamut,Horvat}.  The authors of 
the MQB theory 
\cite{MQB} have assumed that the system only goes to the diamagnetic 
$L$ = 0 state in the ZFC condition so that the PME does not exist in 
ZFC data.  We do not think that this assumption is justified.  Since 
all the $L$ $\geq$ 0 states are the solutions of the Ginzburg-Landau 
equations in mesoscopic samples \cite{MQB}, these states should not 
behave very differently.  For the $L$ = 0 state, it is well known 
that the FC and ZFC 
diamagnetic susceptibilities are the same for superconducting samples 
without open holes.  By analogy, the FC 
and ZFC paramagnetic susceptibilities for samples without open holes should not
have a substantial difference in the case of the $L$ $>$ 0 states.  Because Gibbs energies for all the states are quite close 
near $T_{c}$ (Ref.~\cite{MQB}), the thermal energy should be large 
enough to populate high-$L$ states, leading to the paramagnetic 
Meissner effect.  Because of the metastable nature of these states, 
the paramagnetic Meissner effect will strongly depend on the cooling 
or heating rate and on the field history, particularly if the sample 
never goes to the normal state.  Thus the PME naturally explains the 
hysteresis in Fig.~6 noted earlier.

In Fig.~8a and Fig.~8b we show the ZFC susceptibility for the AD 
prepared sample in 15 Oe and 200 Oe, respectively.  For the 15 Oe ZFC 
measurement, the sample~was cooled from 1170 K to 300 K in ``zero'' field (0.03 Oe) and then the 15 Oe field was 
set at 300 K.  For the 
200 Oe ZFC measurement, the sample was cooled from 1200~K to 310 K in 
``zero'' field (0.03 Oe), and then the 200~Oe
\begin{figure}[htb] 
 \ForceWidth{7cm}
\centerline{\BoxedEPSF{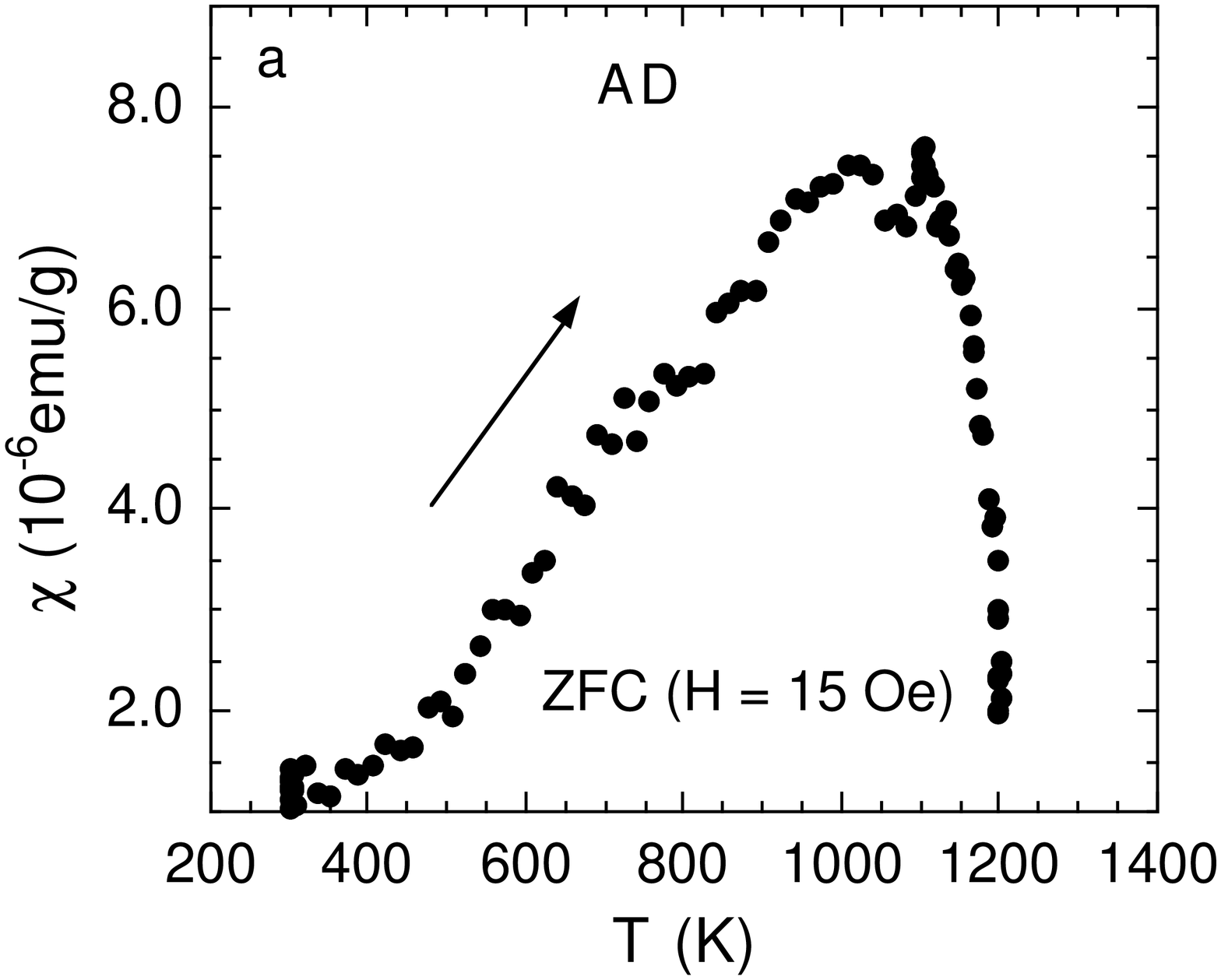}}
\vspace{-0.1cm}
\ForceWidth{7cm}
\centerline{\BoxedEPSF{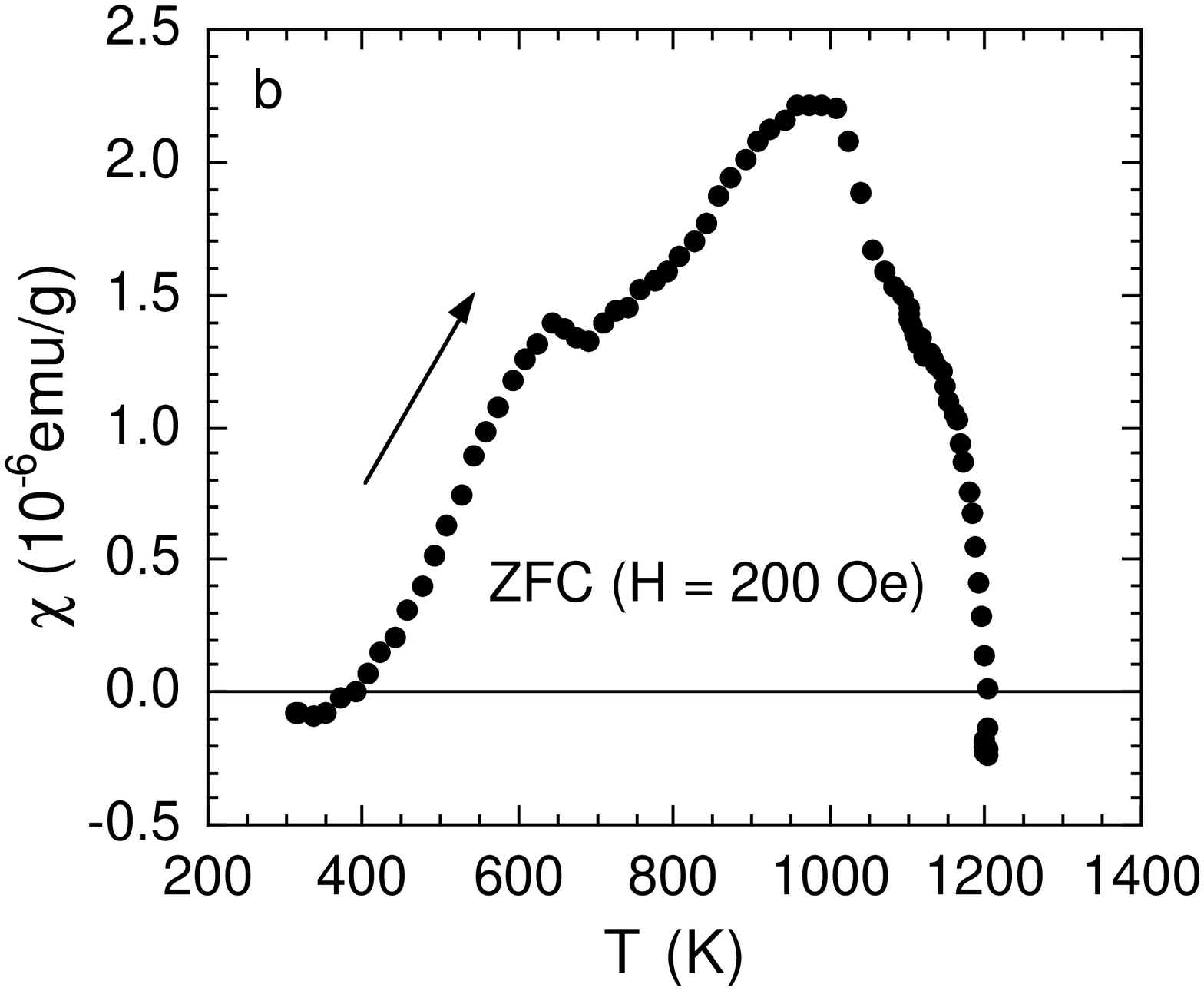}}
	\vspace{0.3cm}
\caption[~]{The temperature dependencies of the ZFC 
susceptibility for the AD prepared MWNT mat sample in magnetic fields 
of a) 15 Oe and b) 200 Oe.  The shoulder and dip features are typical 
of granular superconductors \cite{Wen,Klamut}.  The onset of the 
paramagnetic Meissner effect is at about 1200 K, indicating an 
intragrain superconducting transition temperature of about 1200 K.  }
\end{figure}
\noindent
field was set at 
310 K.  As we learn from the data 
\cite{Klamut} of Ru$_{1-x}$Sr$_{2}$GdCu$_{2+x}$O$_{8-y}$, the onset 
temperature $T_{cJ}$ of intergrain Josephson coupling 
approximately corresponds to either a dip-like or a shoulder-like 
feature.  For the 15 Oe data, there is a shoulder feature near 700 K, 
corresponding to the onset of intergrain Josephson coupling.  A sharp 
peak is seen at 1105 K.  Between 1105 and 1200 K, the susceptibility 
drops sharply with temperature, indicating the onset of the PME at 
about 1200 K.  This implies that the superconducting transition 
temperature is slightly higher than 1200 K because the onset of the 
PME is slightly lower than the intragrain superconducting transition 
temperature ($T_{c}$) \cite{Pust,MQB,Bra,Horvat}.  For the 200 Oe 
data, there is a dip-like feature at about 690 K, corresponding to 
the 
onset of intergrain Josephson coupling \cite{Klamut}.  Between 1000 
and 1200 K, the susceptibility drops 
sharply with temperature and even becomes negative at 1200 K.

In Fig.~9, we show the ZFC susceptibility in four different fields.  
For the 500 Oe and 1000 Oe ZFC measurements, the sample was cooled 
from 1200 K to 310 K in ``zero'' field (about 2 Oe), and the fields were set 
at about 320 K.  For the 15 Oe and 200 Oe ZFC measurements, the 
``zero'' field is 0.03 Oe.   A dip-like feature defines $T_{cJ}$ for the 
200 Oe data while shoulder-like features define $T_{cJ}$ for the 15 Oe, 500 Oe and 1000 
Oe data. 

\begin{figure}[htb] 
 \ForceWidth{7cm}
\centerline{\BoxedEPSF{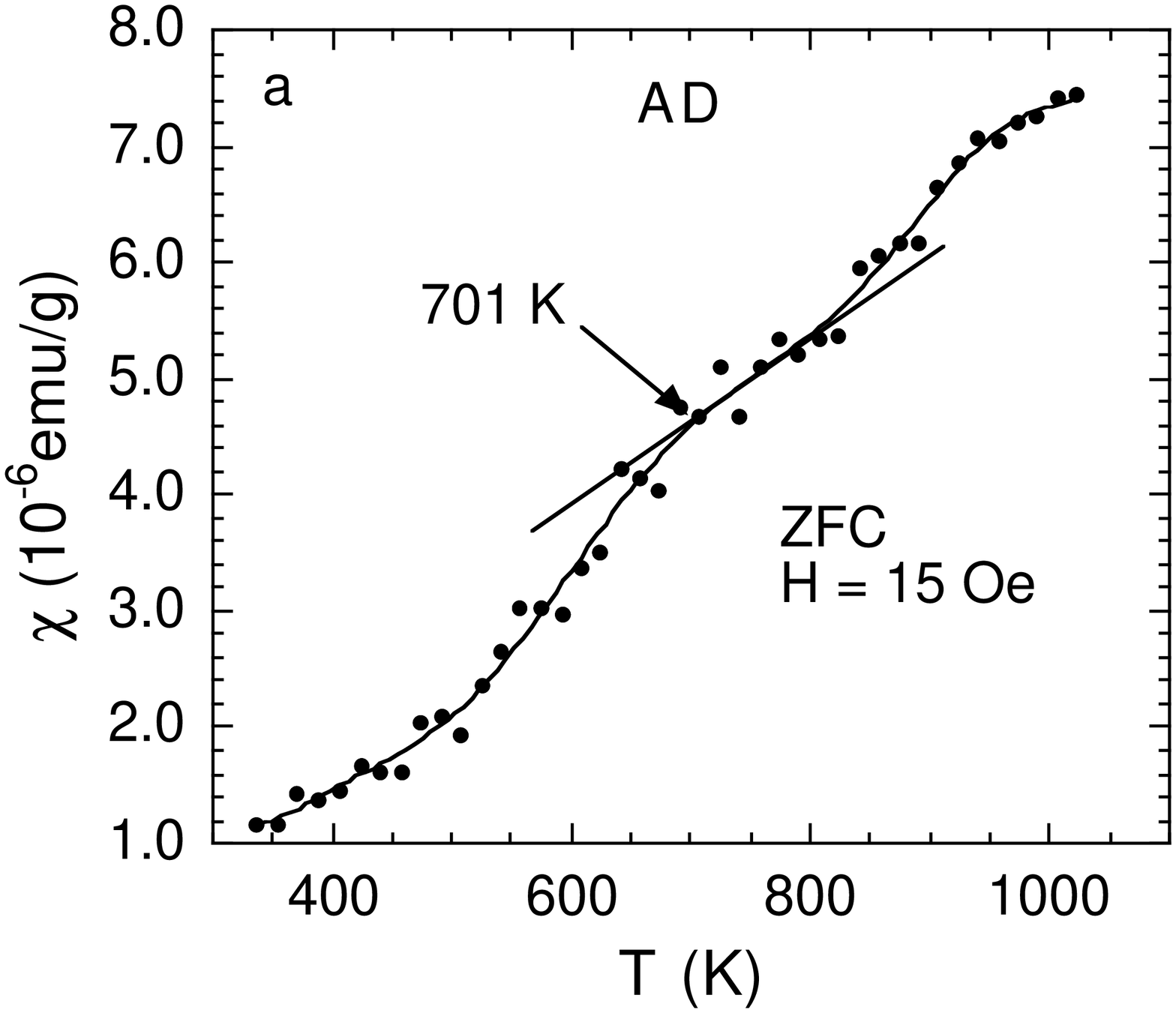}}
\vspace{-0.2cm} 
\ForceWidth{7cm}
\centerline{\BoxedEPSF{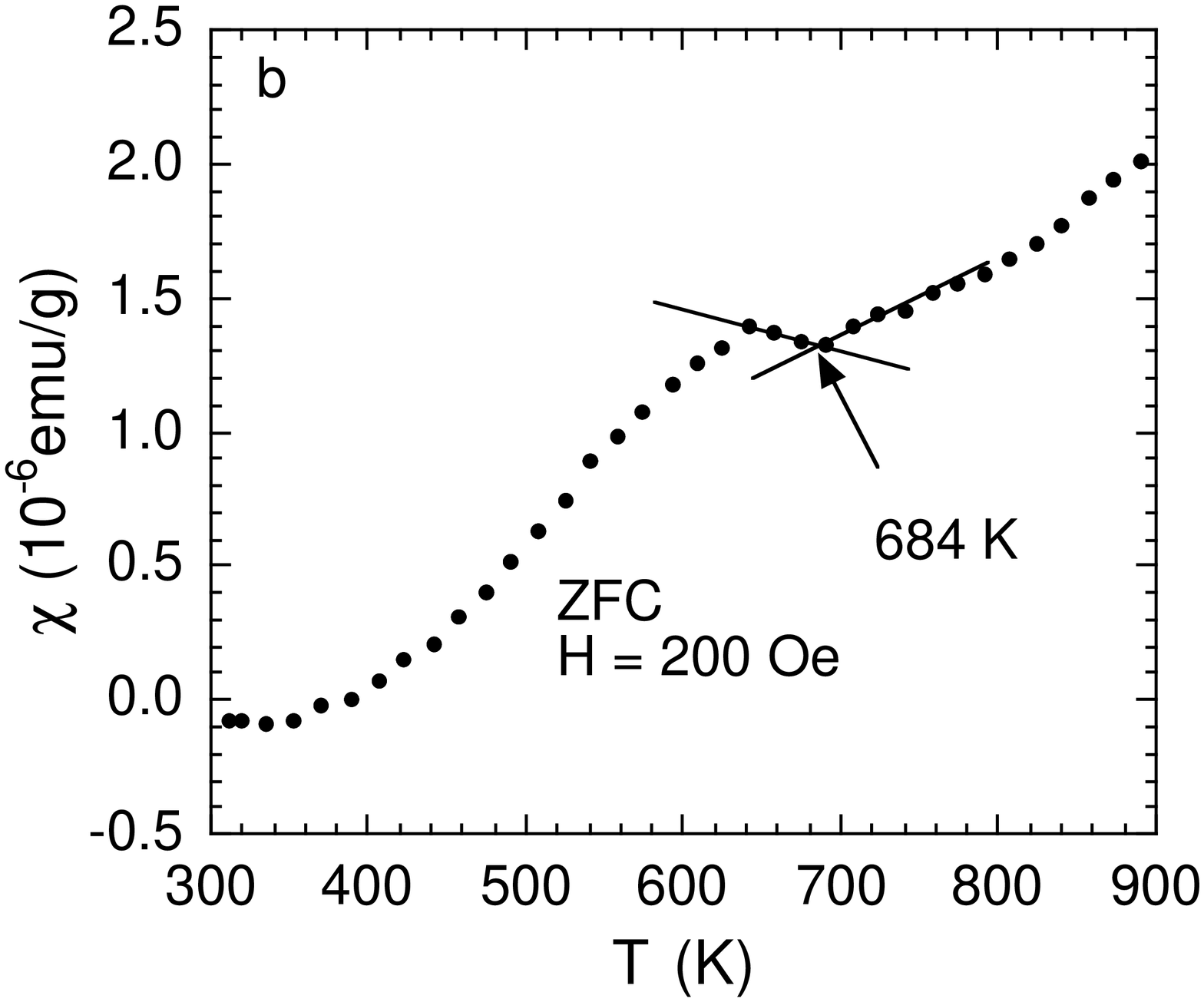}}
\vspace{-0.2cm} 
\ForceWidth{7cm}
\centerline{\BoxedEPSF{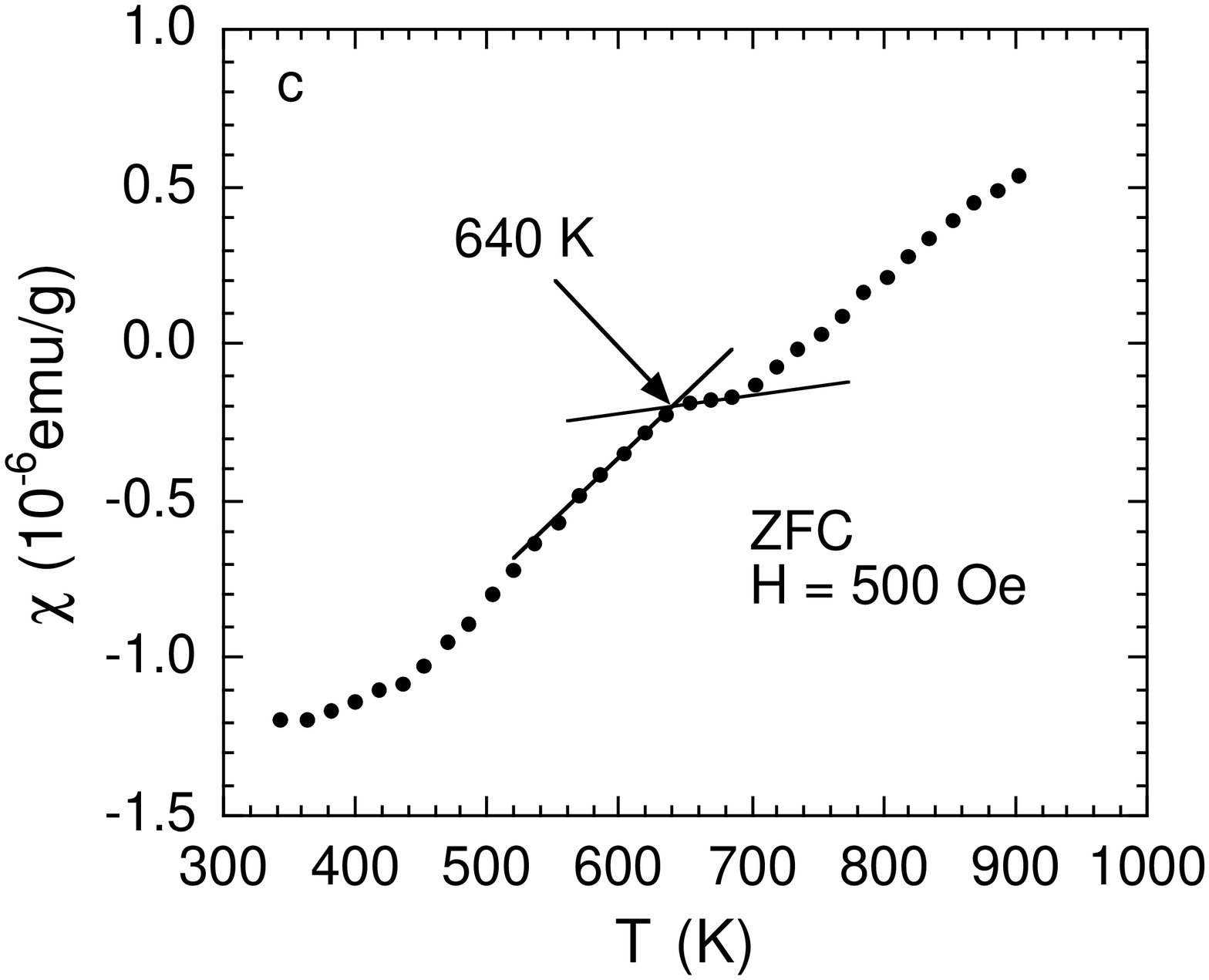}}
\vspace{-0.2cm} 
\ForceWidth{7cm}
\centerline{\BoxedEPSF{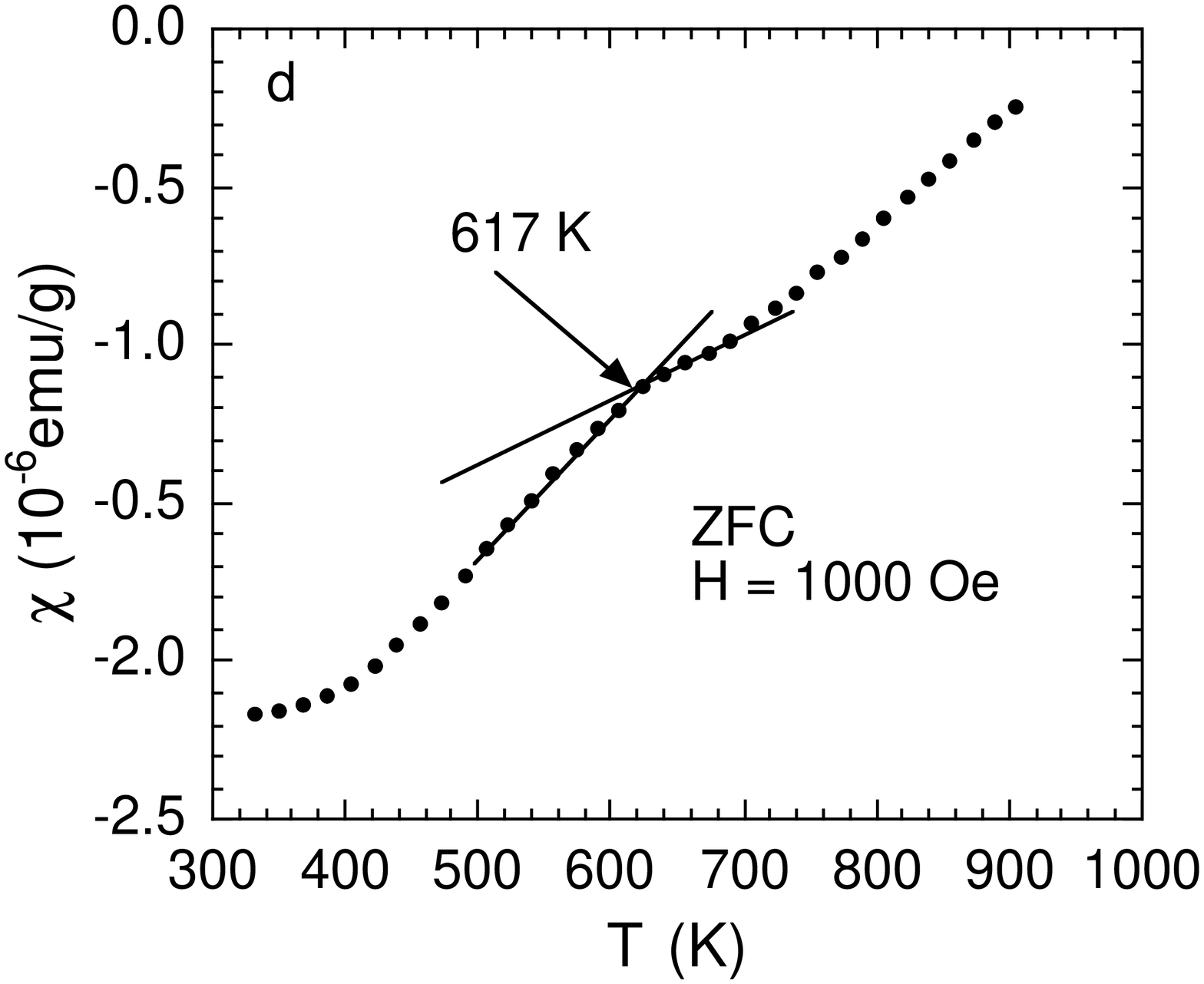}}
 \vspace{0.3cm} 
 \caption[~]{The 
expanded views of the ZFC susceptibility for the AD prepared MWNT mat 
sample in the magnetic fields of a) 15 Oe, b) 200 Oe, c) 500 Oe, and 
d) 1000 Oe. }
\end{figure}

In Fig.~10a, we plot $H$ versus $T_{cJ}$ for the AD prepared MWNT mat 
sample.  
One can see that $T_{cJ}$ depends very strongly on the magnetic field 
in this low field region.  Such a field dependence of 
$T_{cJ}$ is a hallmark of intergrain Josephson coupling 
\cite{Wen,Jardim}.  In Fig.~10b, 
we show $H$ versus $T_{cJ}$ for the granular 
superconductor Ru$_{0.6}$Sr$_{2}$GdCu$_{2.4}$O$_{8-y}$.  $T_{cJ}$ is 
defined as the onset of the paramagnetism in the FC magnetization 
data 
in Fig.~3 of Ref.~\cite{Klamut} and as the midpoint of the resistive 
transition in Fig.~6 of Ref.~\cite{Klamut}.  By comparing Fig.~10a 
with 
Fig.~10b, we can see that the field dependence of $T_{cJ}$ has a 
similar form for both systems, suggesting the same physical origin: 
Intergrain Josephson coupling.  The strong field dependence of 
$T_{cJ}$ in the low field region also rules out a magnetic origin for 
the 
transition.

\begin{figure}[htb] 
 \ForceWidth{7cm}
\centerline{\BoxedEPSF{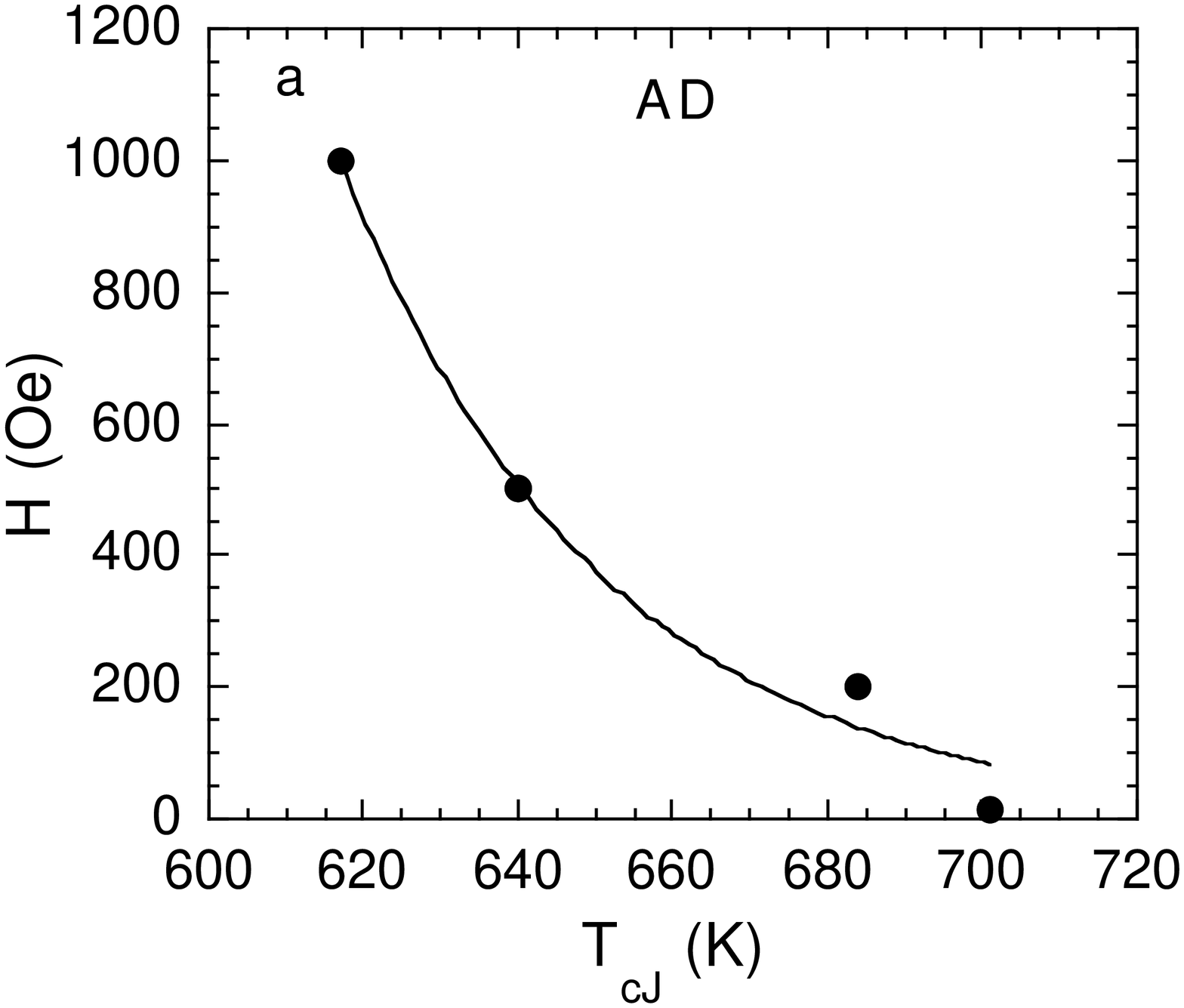}}
\vspace{-0.2cm}
\ForceWidth{7cm}
\centerline{\BoxedEPSF{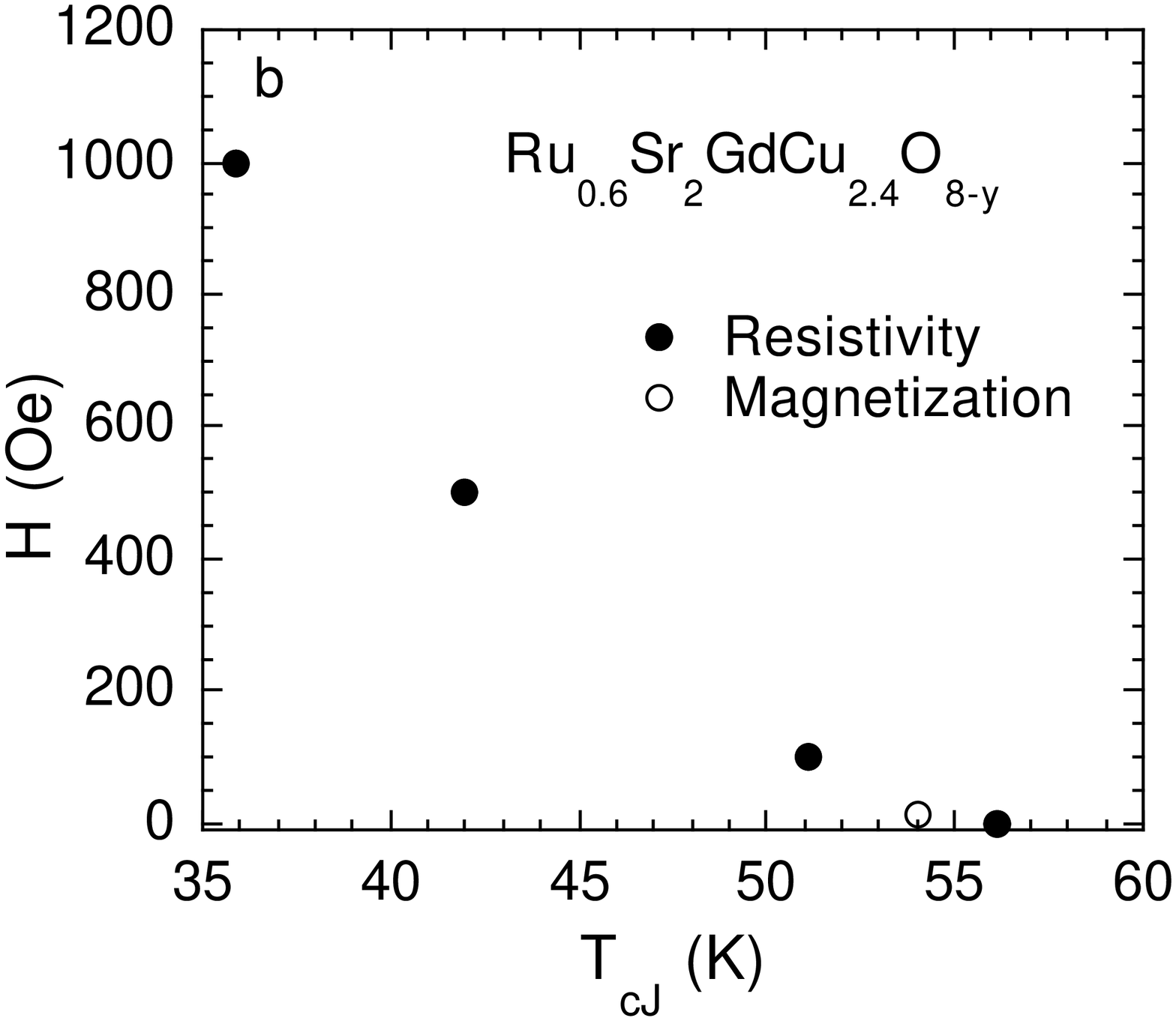}} 
\vspace{0.3cm} 
\caption[~]{The magnetic field ($H$) versus the onset temperature of 
intergrain coupling ($T_{cJ}$) for a) the AD prepared MWNT mat sample 
and b) the granular superconductor 
Ru$_{0.6}$Sr$_{2}$GdCu$_{2.4}$O$_{8-y}$ (Ru-1212). The solid line in 
Fig.~10a is a fit by 
an equation $H = (A/T_{cJ})\exp (-T_{cJ}/T_{\circ})$ with $T_{\circ}$ = 
35.8 K.  This equation has been used to fit the $H$ versus 
$T_{cJ}$ data of a granular superconductor \cite{Wen}. }
\end{figure}

In Fig.~11a we plot previously published resistance data 
for an AD prepared MWNT mat that comes from the same lot as the AD 
prepared sample here.  The 
data are the same as those reported in Ref.~\cite{Zhao1} except they 
are 
thinned here for clarity.  It is interesting that the 
resistance decreases with increasing temperature below 
about 570 K while above~~570 K the resistance tends to 
\begin{figure}[htb] 
 \ForceWidth{7cm}
\centerline{\BoxedEPSF{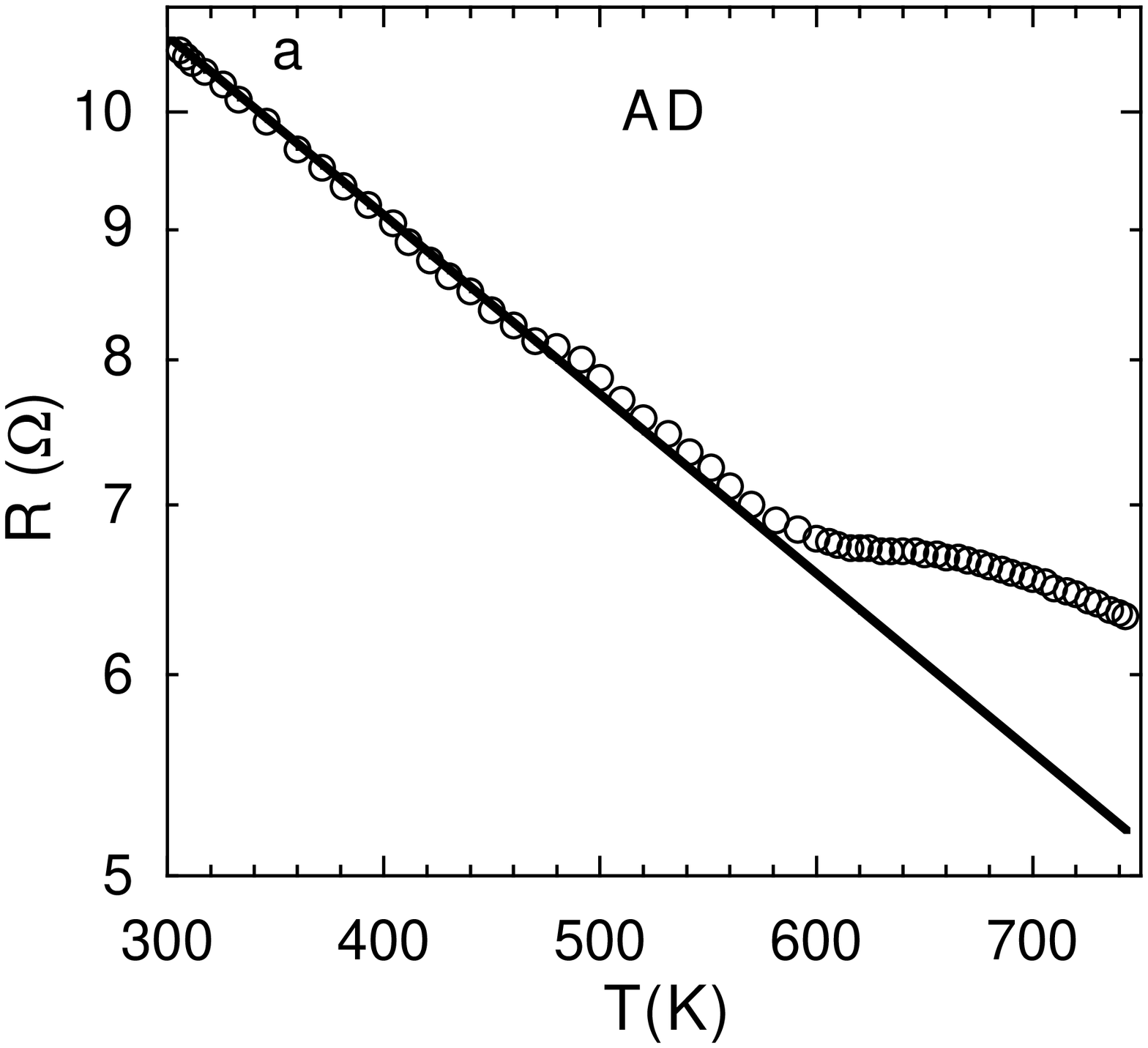}}
\vspace{-0.1cm} 
\ForceWidth{7cm}
\centerline{\BoxedEPSF{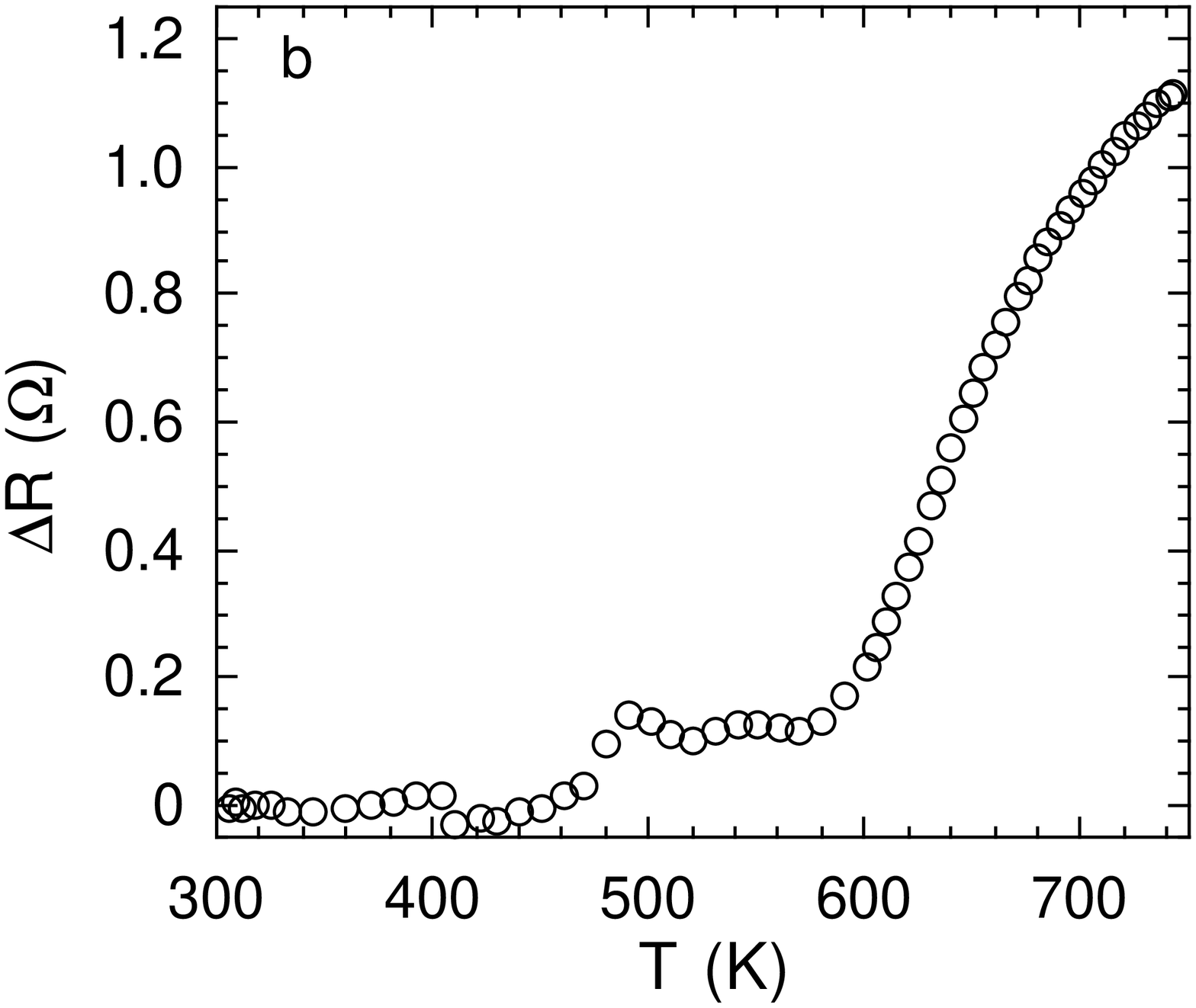}}
\vspace{-0.1cm} 
\ForceWidth{7cm}
\centerline{\BoxedEPSF{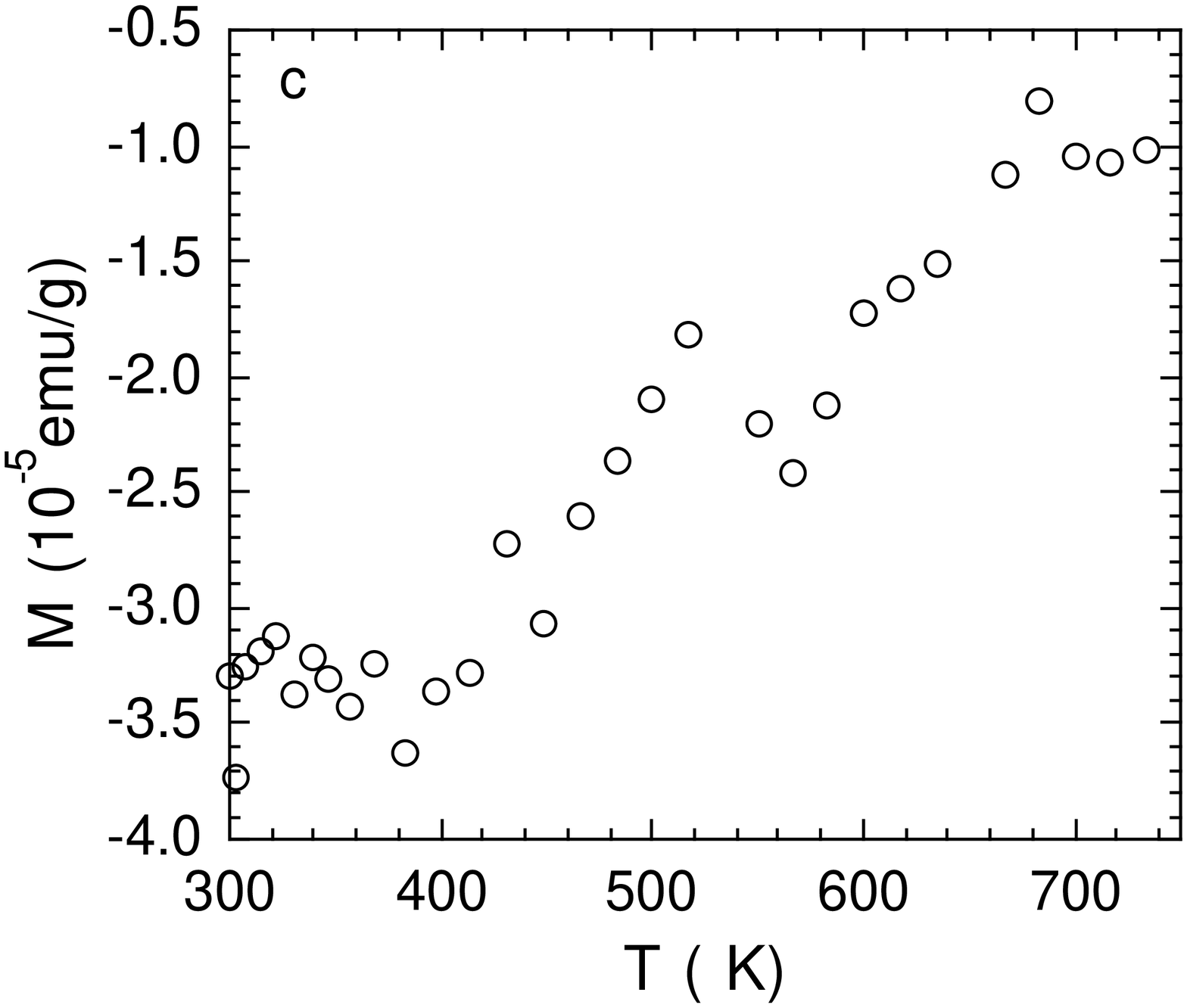}}
	\vspace{0.3cm}
\caption[~]{a) The temperature dependence of the resistance for 
an AD prepared MWNT mat that comes from the same lot as those studied 
here.  The resistance data are the same as those reported in 
Ref.~\cite{Zhao1} except they are thinned for clarity.  A fitted curve 
is given by the solid line.  b) The resistance data subtracted by the fitted 
curve.  c) The expanded view of the warm-up magnetization (Fig.~6) 
for the AD prepared MWNT sample.  The two transitions at about 500 K 
and 700 K in these magnetization data also emerge in the resistance 
data (b) for the same sample.}
\end{figure}
\noindent
turn up.  The 
resistance between 300 K and 450 K can be excellently described by 
17.3$\exp (-T/618.3)$ $\Omega$.  This semiconducting-like resistance 
may come from the intertube barrier resistance for some extremely 
weak 
Josephson junctions in the percolative network.  These Josephson 
junctions are too weakly coupled to lead to a zero-resistance state.  
Instead, the ground state is insulating in these extremely weakly 
coupled junctions.  On the other hand, from Fig.~9 we see that some other 
Josephson junctions in the mat sample are strongly coupled so that the 
onset temperature of intergrain Josephson coupling is around 700 K.  
This suggests that our mat samples consist of both weakly and strongly 
coupled Josephson junctions.

In order to more clearly see the resistive transition, we plot in 
Fig.~11b $\Delta R$, the 
resistance data subtracted by the fitted curve.  These 
$\Delta R$ data should represent the resistive behavior in some parts 
of the mat sample, which have a rather strong Josephson coupling.  In 
addition to the major transition at about 700 K, consistent with the 
$T_{cJ}$ value deduced from the magnetic data above, there appears to 
be a second transition at about 500 K.  This second transition may be 
related to weaker Josephson coupling in some regions of the mat 
sample resulting in this lower $T_{cJ}$.

In Fig.~11c, we show an expanded view of 
the warm-up magnetization below 750 K for the AD prepared MWNT 
sample. We clearly 
see that there is a rapid drop in the magnetization below about 
700 K. In addition to the clear anomaly at 
about 700 K, there is also an anomaly at about 500 K in the 
magnetization data. It is striking that both anomalies are also seen 
in the resistance 
data (Fig.~11b).  This one-to-one correspondence between the 
transitions in the magnetic and electrical data in the same sample 
provides additional evidence for hot superconductivity in AD prepared 
multi-walled carbon nanotubes.

\subsection{Superconductivity in sample CVD1}

Fig.~12a shows the temperature dependence of the warm-up 
magnetization in a 
field of $-$0.06 Oe for sample CVD1.  When the 
sample is inserted into the sample chamber with a field of $-$0.06 
Oe, it 
first experiences a positive field of about 200 Oe and then 
the same negative field due to the presence of the linear 
motor used to vibrate the sample.  Therefore, the warm-up 
magnetization is the sum of the negative remanent magnetization and
the non-remanent magnetization in $-$0.06 Oe.  It is apparent that 
the negative magnetization has a kink feature at about 530 K and that 
the magnetization flattens out above 880 K.  The small constant 
magnetization between 880 and 1000 K (represented by the horizontal 
line) in the warm-up data is the same as that in the cool-down data.  
Thus, the reversible constant magnetization between 880 and 1000 K is 
the nonremanent magnetization, which suggests a 
ferrimagnetic/ferromagnetic ordering at about 880 K.  From the Curie 
temperature of 880 K, we conclude that this CVD prepared sample also 
contains Fe$_{3}$O$_{4}$ magnetic impurities.

Assuming this small constant nonremanent magnetization extends to the 
temperature region between 300 and 880 K, we obtain the remanent 
magnetization by subtracting this small constant term from the total 
magnetization, as shown in Fig.~12b.  The remanence above 530 K can 
be well 
fitted by an equation
\begin{equation}
M_{r}(T) = M_{r}(0)[1- (T/T_{C})^{q}].
\end{equation}
The solid line is the fitted curve with a fixed parameter $T_{C}$ = 
880 
K and two fitting parameters: $q$ = 1.5 and $M_{r}(0)$ = $-$0.0024 
emu/g.  
The value of $q$ = 1.5 is typical for multidomain ferromagnetic 
particles.

\begin{figure}[htb] 
 \ForceWidth{7cm}
\centerline{\BoxedEPSF{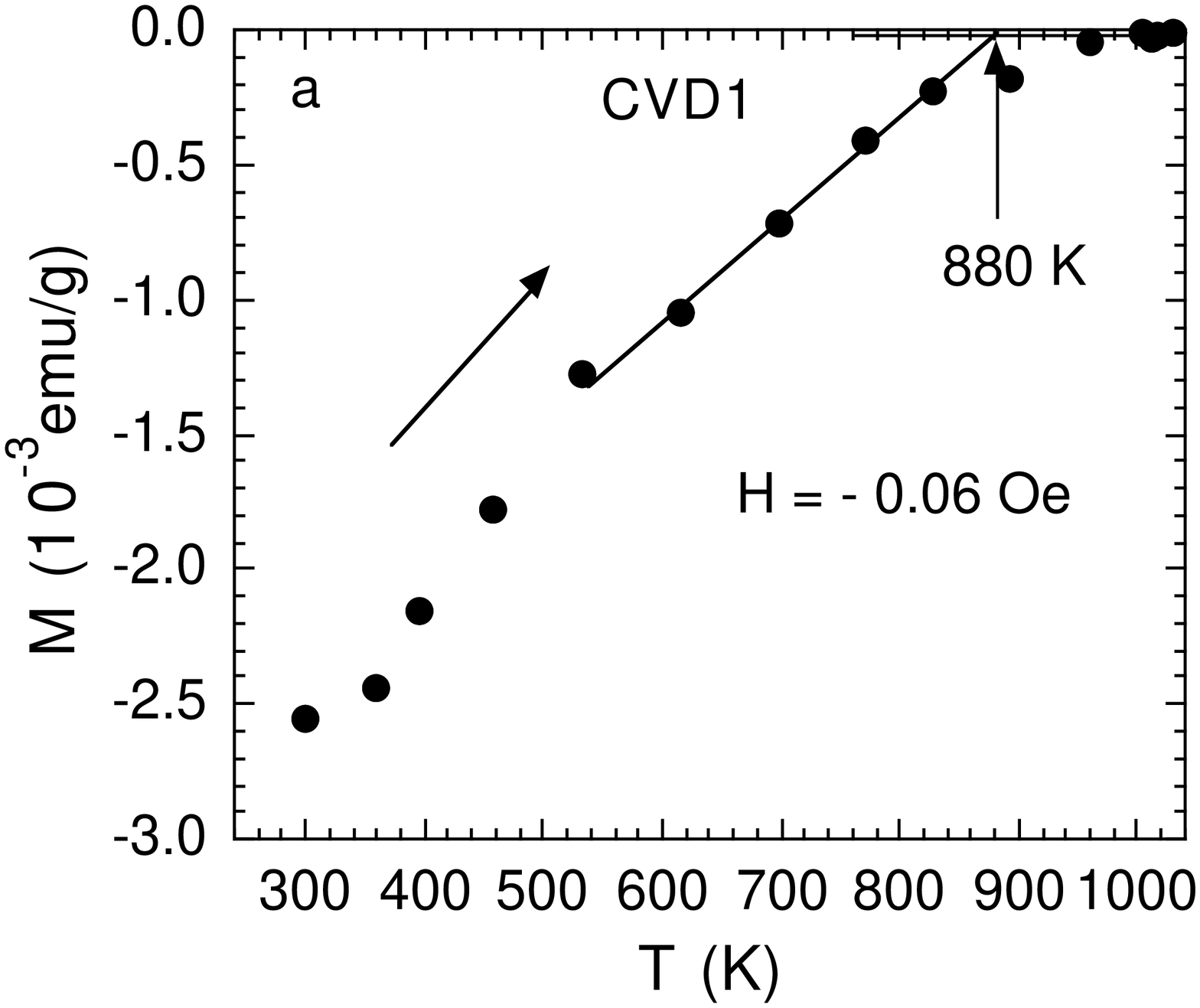}}
\vspace{-0.1cm} 
\ForceWidth{7cm}
\centerline{\BoxedEPSF{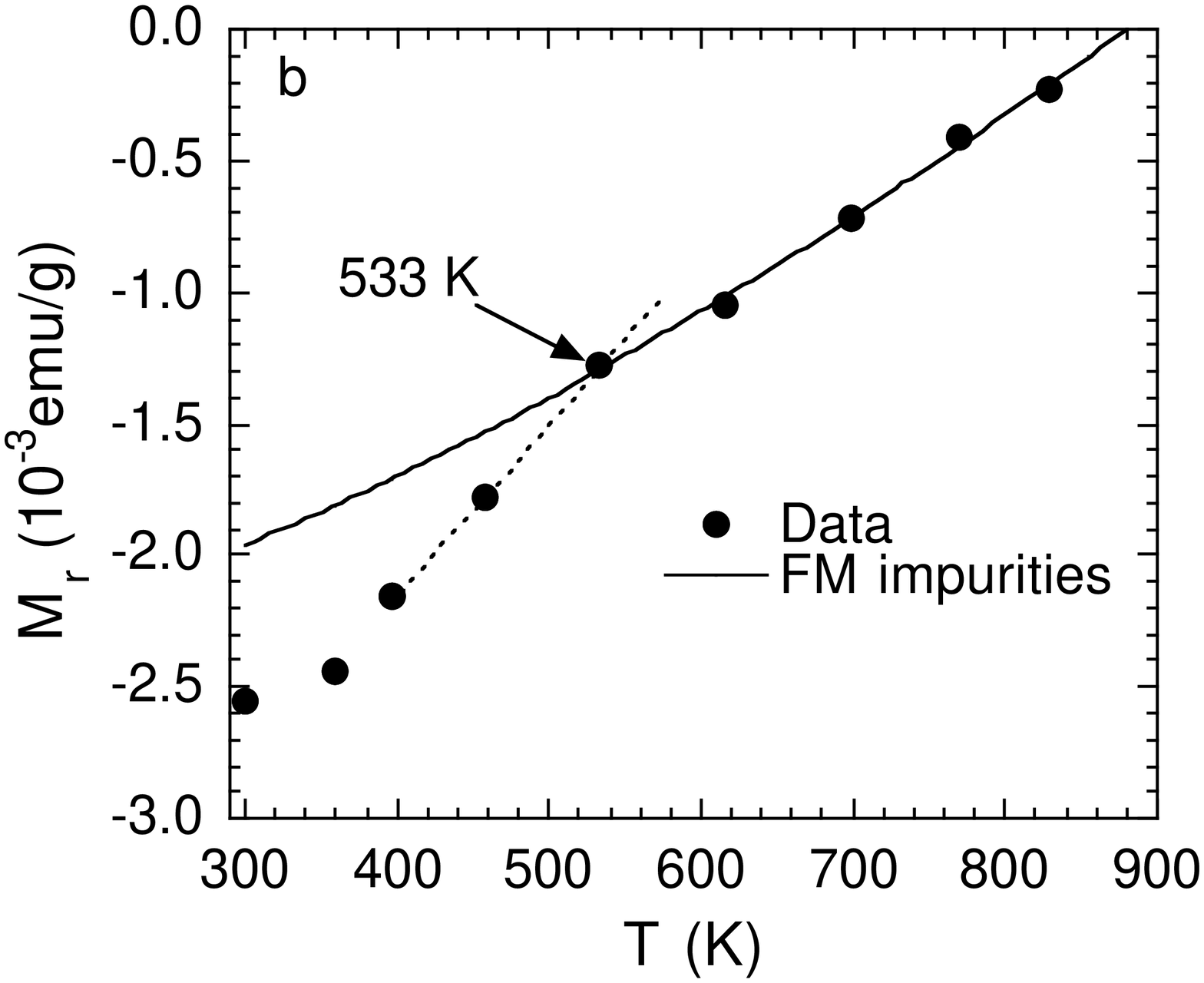}}
\vspace{0.3cm}
\caption[~]{a) The temperature dependence of the warm-up magnetization 
in a field of $-$0.06 Oe for sample CVD1.  b) The temperature dependence of the remanent magnetization 
for sample CVD1.  The solid line represents the ferrimagnetic 
remanence due to the Fe$_{3}$O$_{4}$ magnetic impurities. }
\end{figure}

From Fig.~12b, we  see that an extra remanence sets in below 
about 533 K.  This extra contribution should be 
associated with a second transition which could be the onset of 
intergrain Josephson coupling or a ferromagnetic/ferrimagnetic 
transition.  We can make a clear distinction between these two 
interpretations by studying the field dependence of this transition.  
If this transition very strongly depends on the magnetic field, it 
must be related to the onset temperature of intergrain Josephson 
coupling.

Fig.~13a shows the temperature dependencies of the FC susceptibility 
for 
sample CVD1 with different magnetic fields.  One can see that there 
is 
a ferromagnetic/ferrimagnetic phase transition with the Curie 
temperature 
$T_{C}$ of about 870 K.  This is consistent with  the remanence data in Fig.~12b.

\begin{figure}[htb] 
 \ForceWidth{7cm}
\centerline{\BoxedEPSF{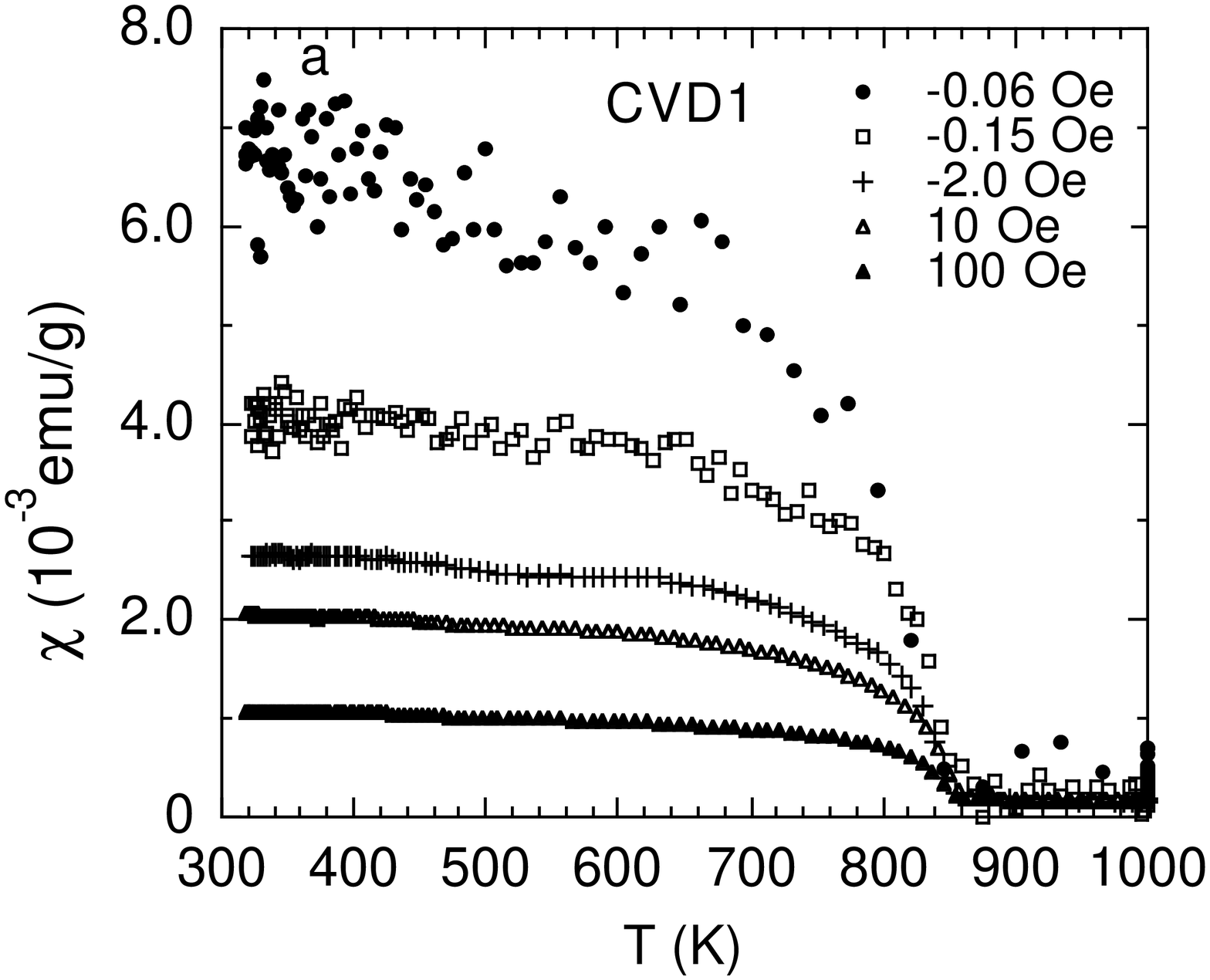}}
\vspace{-0.1cm}
\ForceWidth{7cm}
\centerline{\BoxedEPSF{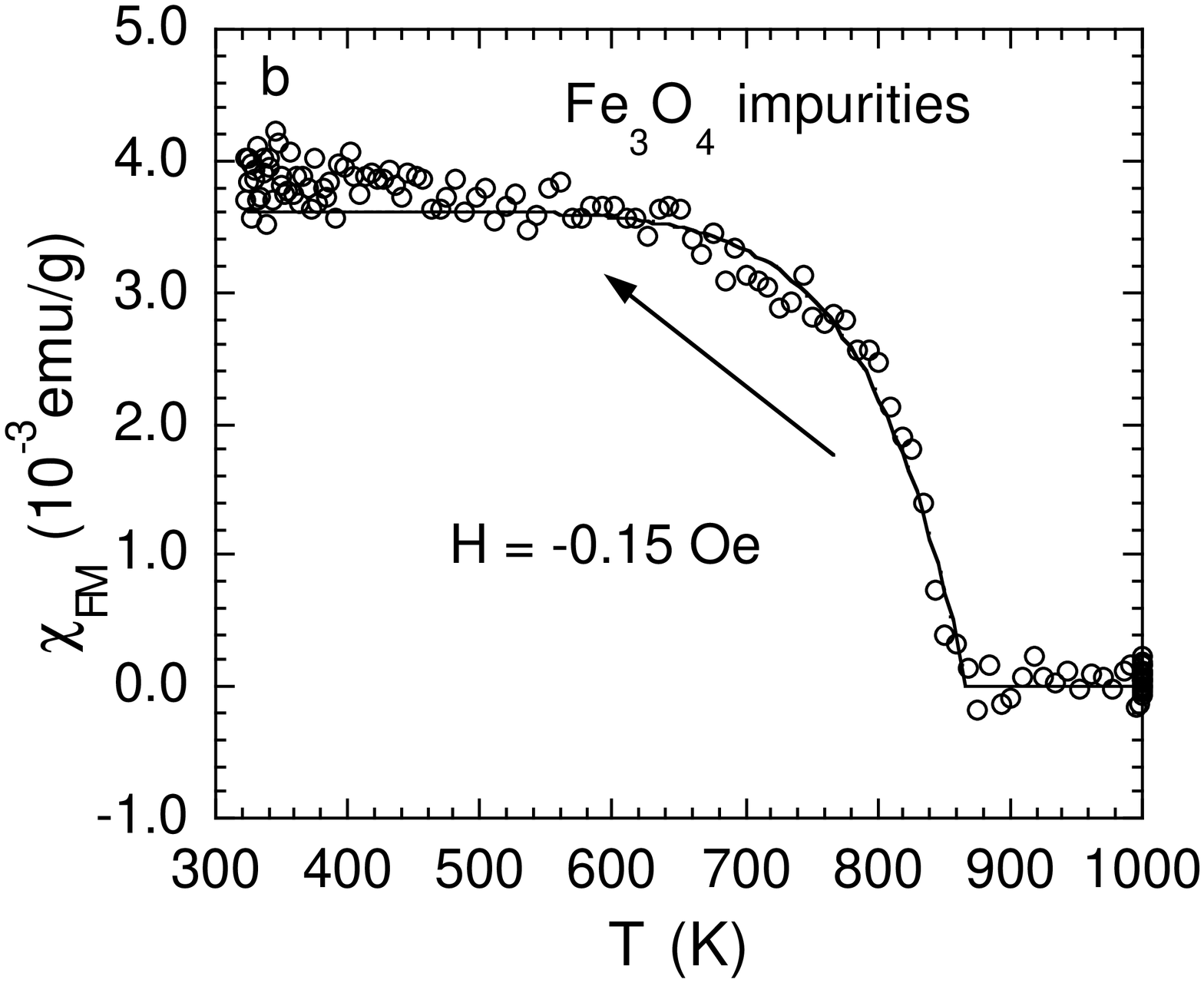}}
\vspace{-0.1cm}
\ForceWidth{7cm}
\centerline{\BoxedEPSF{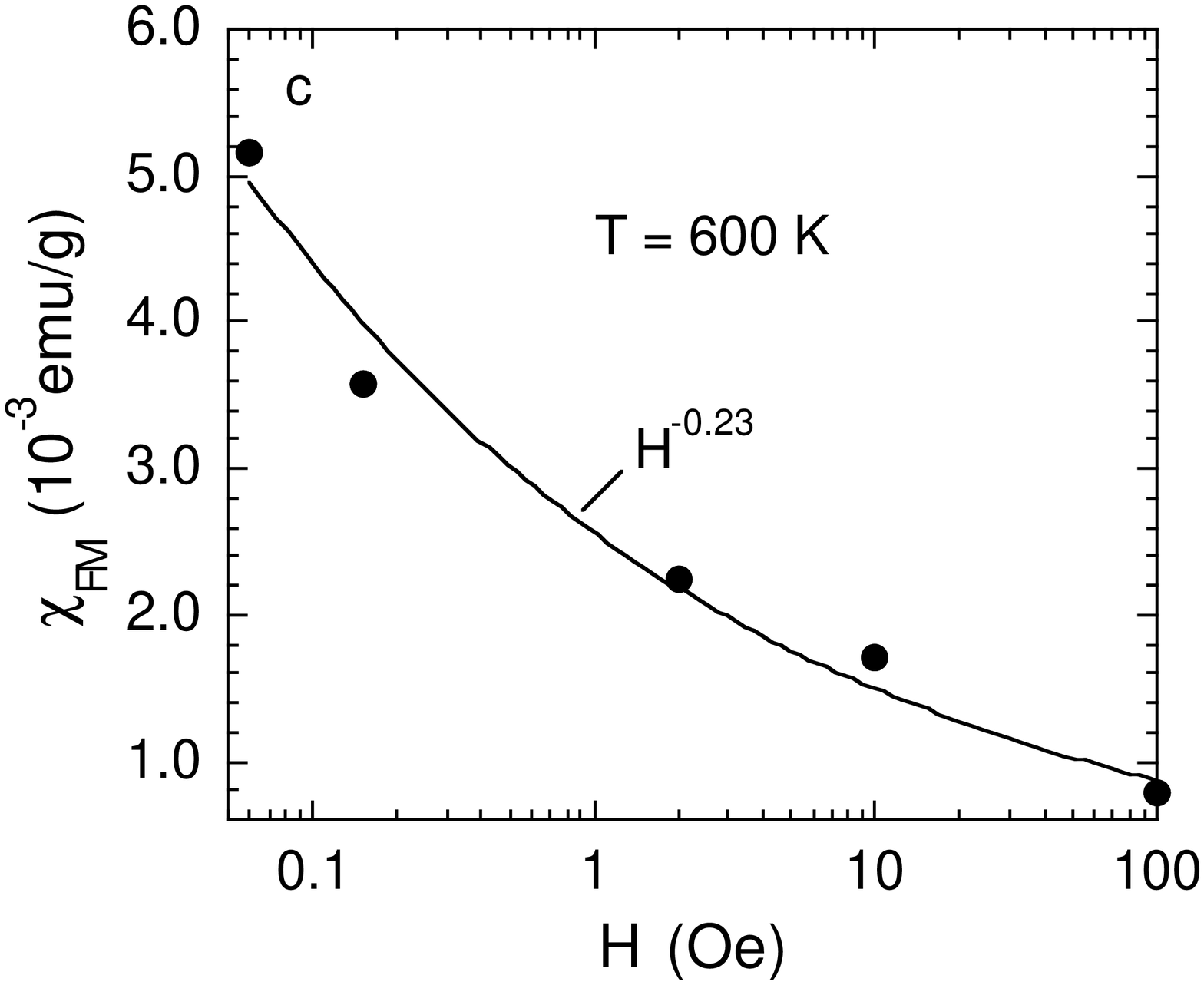}}
\ForceWidth{7cm}
\vspace{0.3cm}
\caption[~]{a) The temperature dependence of the FC susceptibility for 
sample CVD1 with different magnetic fields.  b) The temperature 
dependence of the ferrimagnetic 
component of the susceptibility in the field of $-$0.15 Oe.  c) The 
ferrimagnetic component of the susceptibility at 600 K versus $H$ for 
sample CVD1.  }
\end{figure}

In Fig.~13b we plot the temperature dependence of the ferrimagnetic 
component of the susceptibity in the field of $-$0.15 Oe.  The 
ferrimagnetic
component is estimated by subtracting a small constant paramagnetic 
term$-$evident between $T_{C}$ and 1000 K$-$from the total susceptibility.  The 
solid 
line is a fit by Eq.~2 with the fitting parameters: $T_{C}$ = 867.6 K and 
$p$ = 11.66.  This $p$ value is very close to that for the Fe 
impurities in the AD sample (see Fig.~7a).  In Fig.~13c, we plot the 
ferrimagnetic component of the susceptibility at 600 K versus $H$ for 
sample CVD1.  It is clear that the ferrimagnetic susceptibility 
decreases with increasing $H$, which follows a power law: $\chi_{FM}$ 
$\propto$ $H^{-0.23}$.

This unusual field dependence is hard to explain if there were no 
superconductivity in the nanotubes. From the measured magnetic 
hysteresis loop 
at 300 K, we obtain the saturation magnetization of 1.4 emu/g.  Since 
the saturation magnetization of pure Fe$_{3}$O$_{4}$ particles is 92 emu/g at room temperature, we estimate that the sample has 1.5$\%$
Fe$_{3}$O$_{4}$ magnetic impurities, which would lead to a low-field 
susceptibility of 0.015$\times$0.045 emu/g = 6.75$\times$10$^{-4}$ 
emu/g.  From Fig.~13c, we see that the ferrimagnetic susceptibility in 
0.06 Oe is 0.0052 emu/g, which is about one order of magnitude larger 
than our predicted value (6.75$\times$10$^{-4}$ emu/g).  On the other 
hand, the estimated ferrimagnetic susceptibility in 100 Oe is only 
18$\%$
larger than the predicted value.  If superconductivity exists in the 
nanotubes, some fraction of the magnetic impurities will be the flux 
pinning centers where the effective field is close to $H_{c1}$.  
Within this picture, the ferrimagnetic susceptibility will be greatly 
enhanced in the field region where $H_{c1}/H$ $>$$>$ 1 while the 
enhancement vanishes for $H$ $>$ $H_{c1}$.  Then, if 0.6$\%$ of magnetic 
impurities are
the pinning centers, the ferrimagnetic susceptibility in the field of 
0.06 Oe will increase by one order of magnitude.  The percentage of 
magnetic impurities being the pinning centers should strongly depend 
on the field and thermal history.  Fluxes can be effectively trapped 
if the sample is cooled in a field above $T_{c}$.  This implies 
that~the percentage is larger when the sample is cooled from a higher 
temperature.  It is also known that more fluxes will be trapped at 
higher fields.  Thus when the field increases the percentage of the 
impurities being pinning centers increases.  However, if this 
percentage increasing rate is lower than the decreasing rate of 
$H_{c1}/H$, the ferrimagnetic susceptibility will decrease with 
increasing field.  This may explain the field dependence of the 
ferrimagnetic susceptibility in sample CVD1 (Fig.~13c).

In Fig.~14, we show the expanded views of the FC 
susceptibility for sample CVD1 with different fields.  In this 
temperature region, the susceptibility of the ferrimagnetic impurities 
increases weakly with decreasing temperature (a negative slope).  
There is an additional onset of paramagnetism, which is clearly seen in 
all the figures.  This local onset of paramagnetism could be 
consistent with ferromagnetic/ferromagnetic ordering or the onset of 
intergrain Josephson coupling \cite{Klamut}.  As one can clearly see 
from Fig.~14, the onset temperature depends very strongly on the 
magnetic field.  Therefore this transition is {\em only} consistent with the 
onset of intergrain Josephson coupling.

We define the crossing temperature of the two straight lines in Fig.~14 to be 
$T_{cJ}$.  In Fig.~15, we show $H$ versus $T_{cJ}$ for sample CVD1.  
It is striking that $T_{cJ}$ depends~very strongly on the field in the 
low field region, but becomes weakly field dependent in higher fields.  
This behavior is 
\begin{figure}[htb] 
\ForceWidth{7cm}
\centerline{\BoxedEPSF{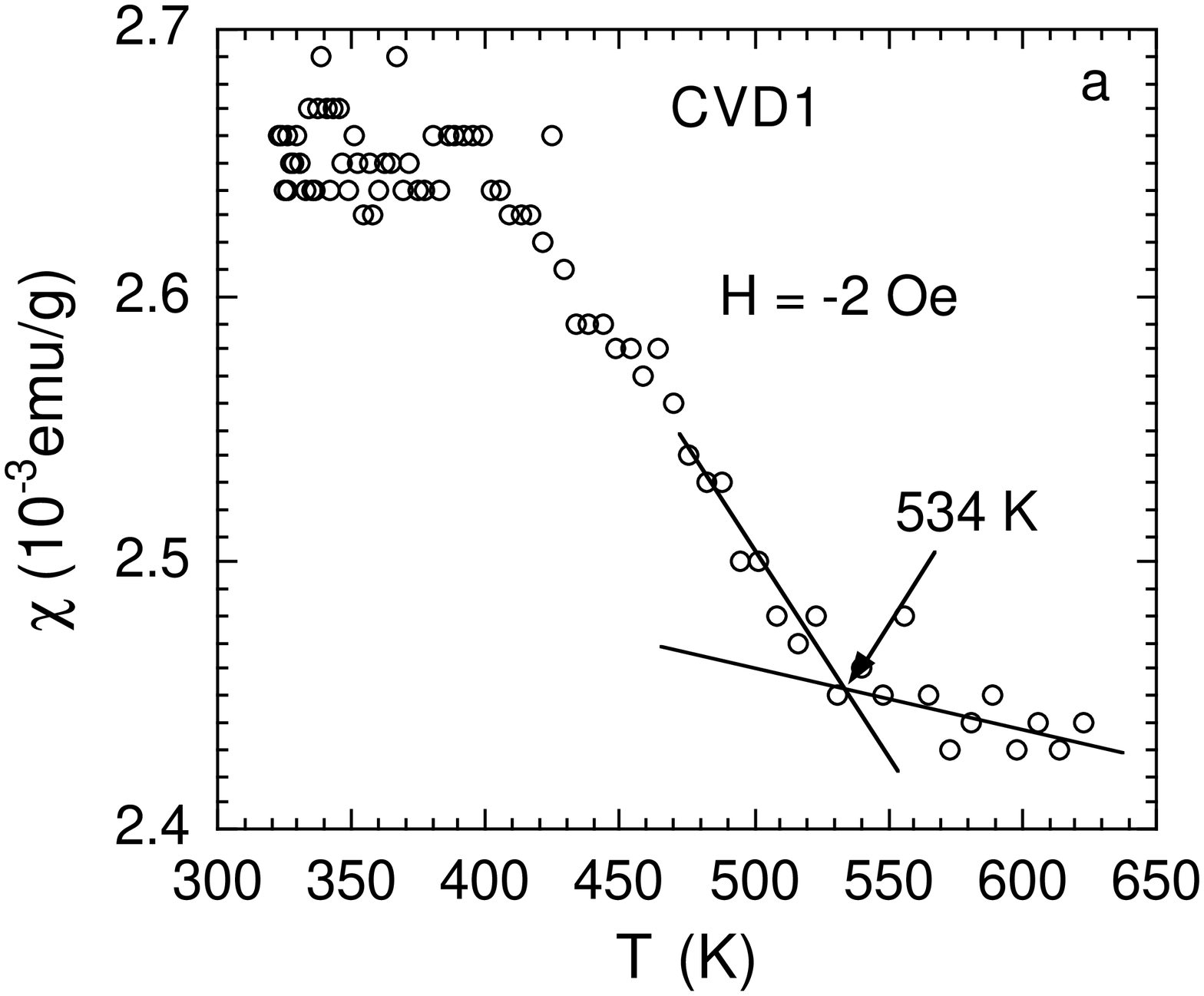}}
\vspace{-0.2cm} 
\ForceWidth{7cm}
\centerline{\BoxedEPSF{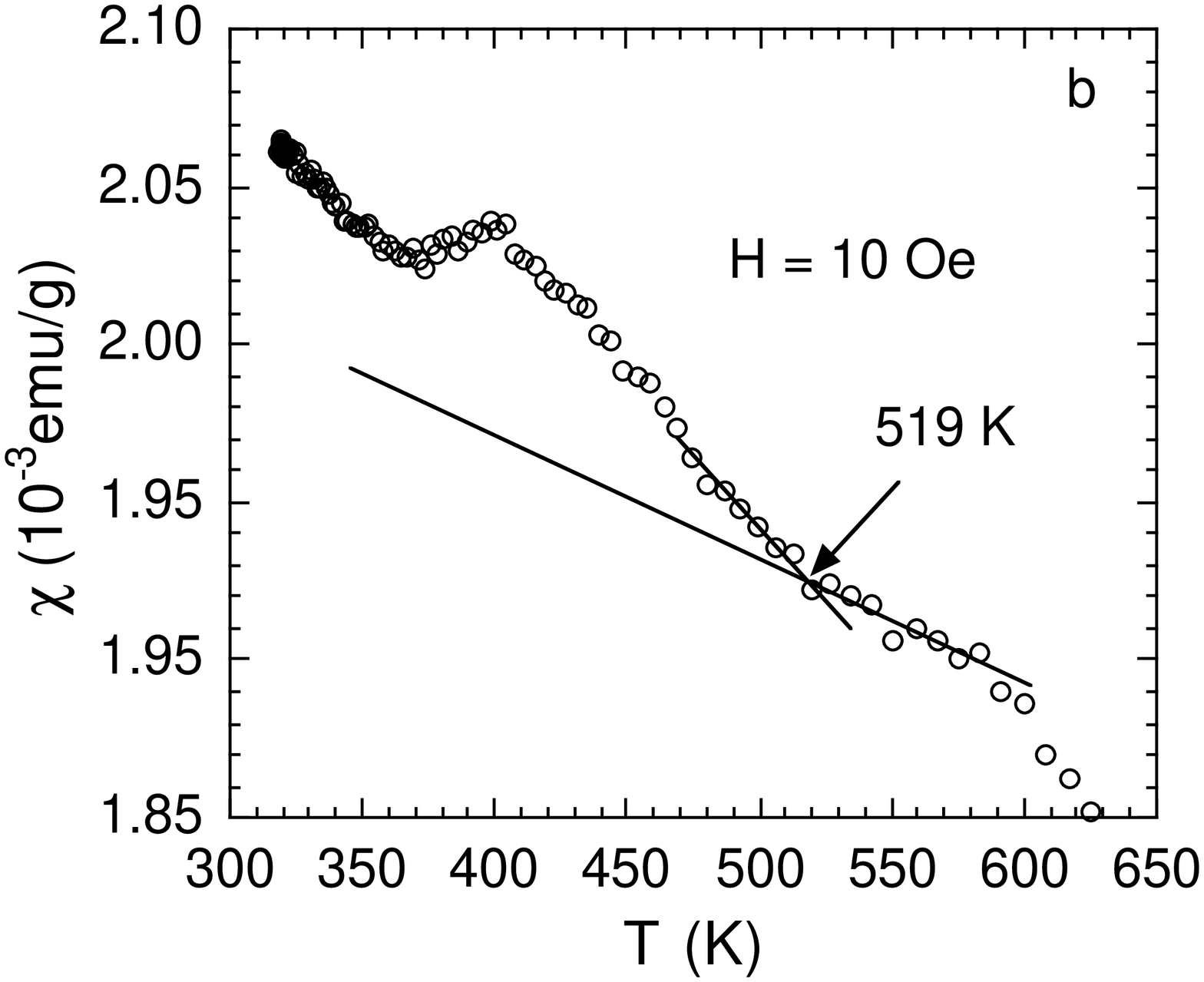}}
\vspace{-0.2cm} 
\ForceWidth{7cm}
\centerline{\BoxedEPSF{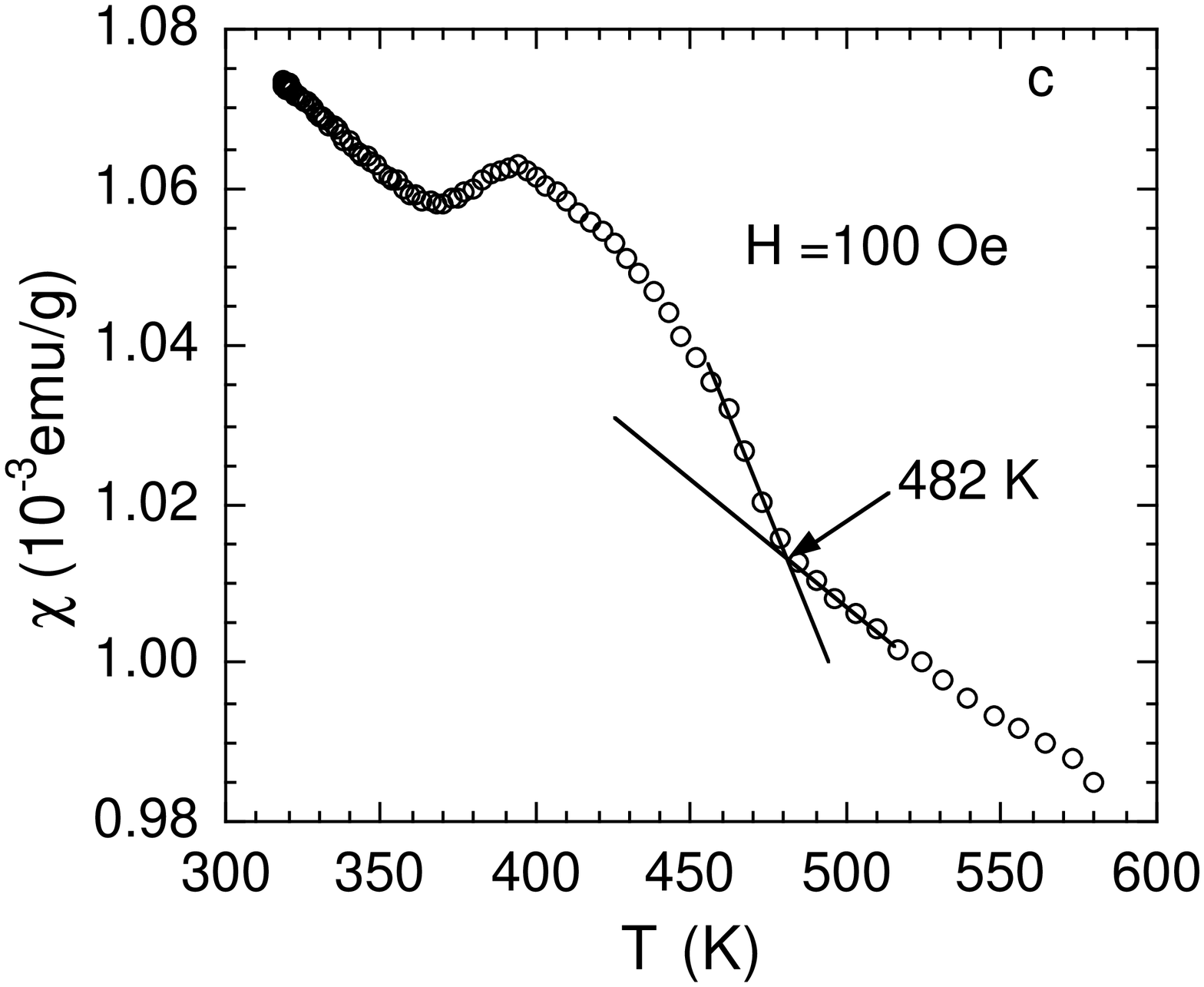}}
\vspace{-0.2cm} 
\ForceWidth{7cm}
\centerline{\BoxedEPSF{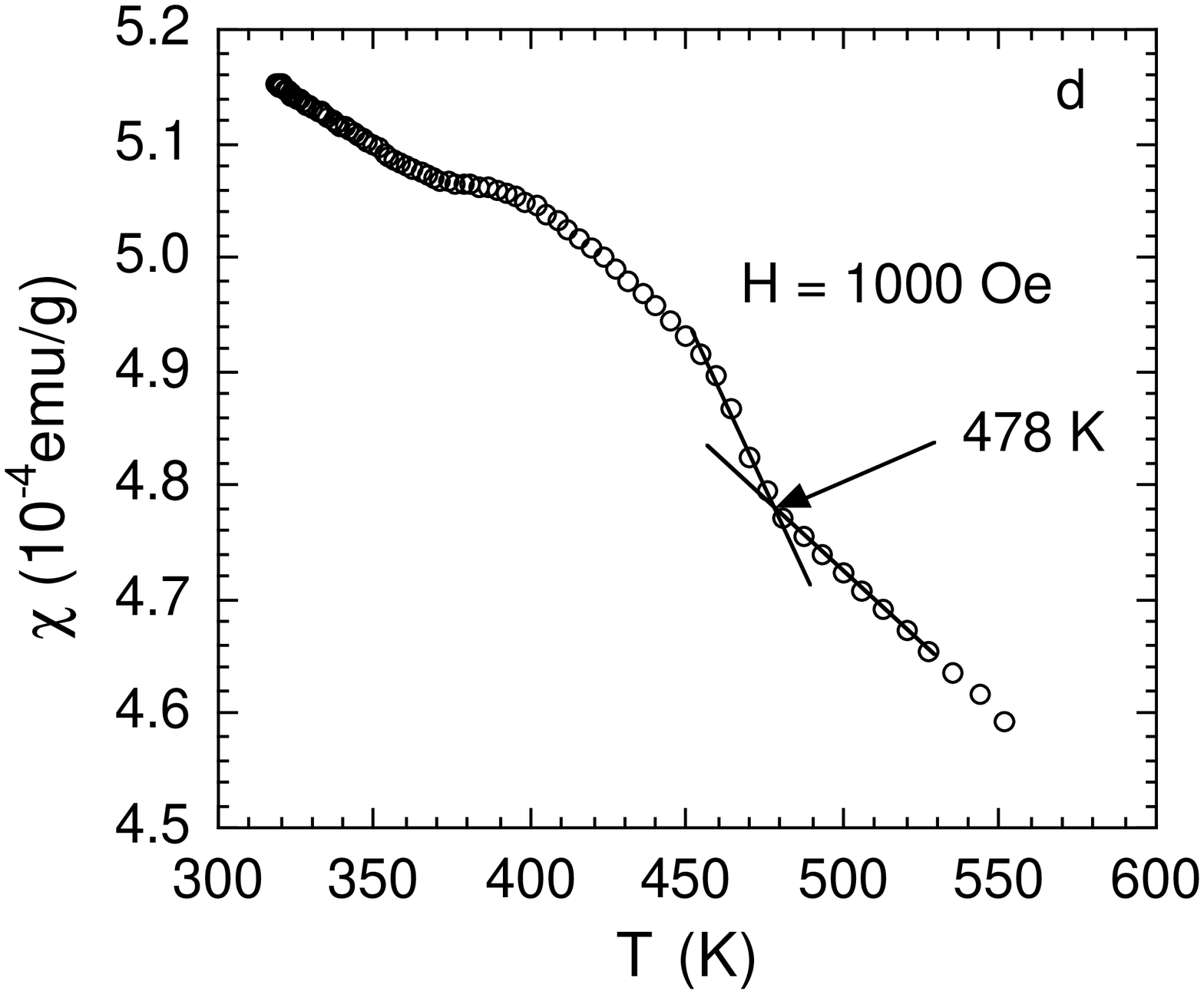}}
 \vspace{0.3cm} 
  \caption[~]{The 
expanded views of the FC susceptibility for sample CVD1 in the magnetic 
fields of a) 2 Oe, b) 10 Oe, c) 100 Oe, and d) 1000 Oe.  }
\end{figure}
\noindent
very similar to that observed in granular 
superconductors \cite{Wen,Klamut,Jardim}.  The strong field dependence 
of $T_{cJ}$ rules out a magnetic origin of this transition.  Further, 
there is a drop in the susceptibility about 100~K below $T_{cJ}$.  This 
feature is also consistent with that observed in granular 
superconductors \cite{Wen,Klamut}.

\begin{figure}[htb] 
 \ForceWidth{7cm}
\centerline{\BoxedEPSF{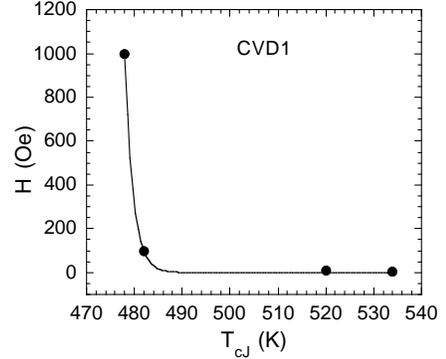}}
	\vspace{0.4cm}
\caption[~]{$H$ versus 
$T_{cJ}$ for sample CVD1.  The solid line is a fit by 
an equation $H = (A/T_{cJ})\exp (-T_{cJ}/T_{\circ})$ with $T_{\circ}$ = 1.75 
K.  }
\end{figure}

Another important signature of granular superconductivity is the 
disappearance of the superconducting remanence above $T_{cJ}$ (see 
Ref.~\cite{Wen,Klamut}).  As shown in Fig.~12b, the ferrimagnetic 
remanence dominates above 533 K in this CVD sample, that is, the 
superconducting remanence is substantial only below 533 K and nearly 
disappears above 533 K.  Thus the remanence data provide evidence for 
$T_{cJ}$ = 533 K in 0.06 Oe.  This is in excellent agreement with that 
determined from the susceptibility data (see Fig.~15).

\subsection{Superconductivity in sample CVD2}

Fig.~16a shows a zoomed view of the warm-up 
magnetization in a field of 0.08 Oe for sample CVD2.  The virgin 
sample 
is inserted into the sample chamber with a field of 0.08 Oe after it 
first 
experiences a positive field of about 200 Oe and then the same 
negative field.  So the warm-up magnetization is the sum of the 
negative 
remanent magnetization and the non-remanent magnetization~in 0.08 Oe.  It is remarkable that the negative magnetization has a kink 
feature at about 700 K and that the magnetization flattens out above 
875 K.  The constant magnetization between 875 and 1000 K (represented 
by the horizontal line) is the nonremanent magnetization in this 
temperature region.  These data imply a ferrimagnetic/ferromagnetic 
ordering at about 875 K.  From the Curie temperature, we conclude that 
this CVD prepared sample also contains Fe$_{3}$O$_{4}$ magnetic 
impurities.  Since the total impurity concentration in this sample is 
about 0.3$\%$, the concentration of
Fe$_{3}$O$_{4}$ magnetic impurities
should be less than 0.3$\%$.

Assuming this small constant nonremanent magnetization extends to the 
temperature region between 300 and 875 K, we obtain the remanent 
magnetization by subtracting this small constant term from the total 
magnetization, as shown in Fig.~16b.  The remanence above 700 K is fitted by 
Eq.~3.  The solid line is the fitted curve with two fixed parameters 
$T_{C}$ = 875 K and $q$ = 1.5, and one fitting parameter $M_{r}(0)$ = 
3.07$\times$10$^{-4}$ emu/g.  The fitted solid line represents the 
ferrimagnetic remanence contributed from the Fe$_{3}$O$_{4}$ 
impurities.  The large extra contribution below 700 K is associated 
with a second transition which should be the onset of intergrain 
Josephson coupling.

\begin{figure}[htb] 
 \ForceWidth{7cm}
\centerline{\BoxedEPSF{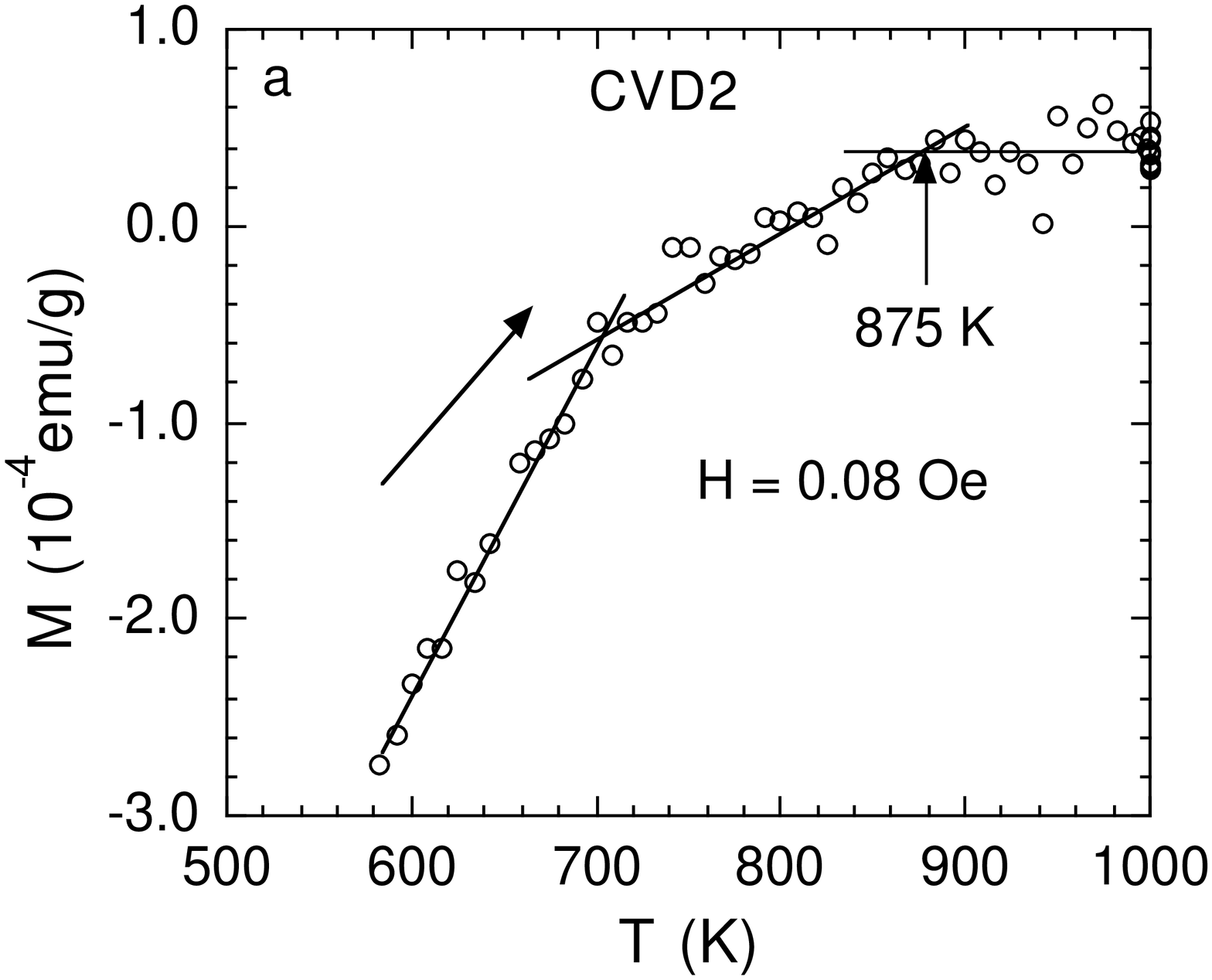}}
\vspace{-0.1cm}
\ForceWidth{7cm}
\centerline{\BoxedEPSF{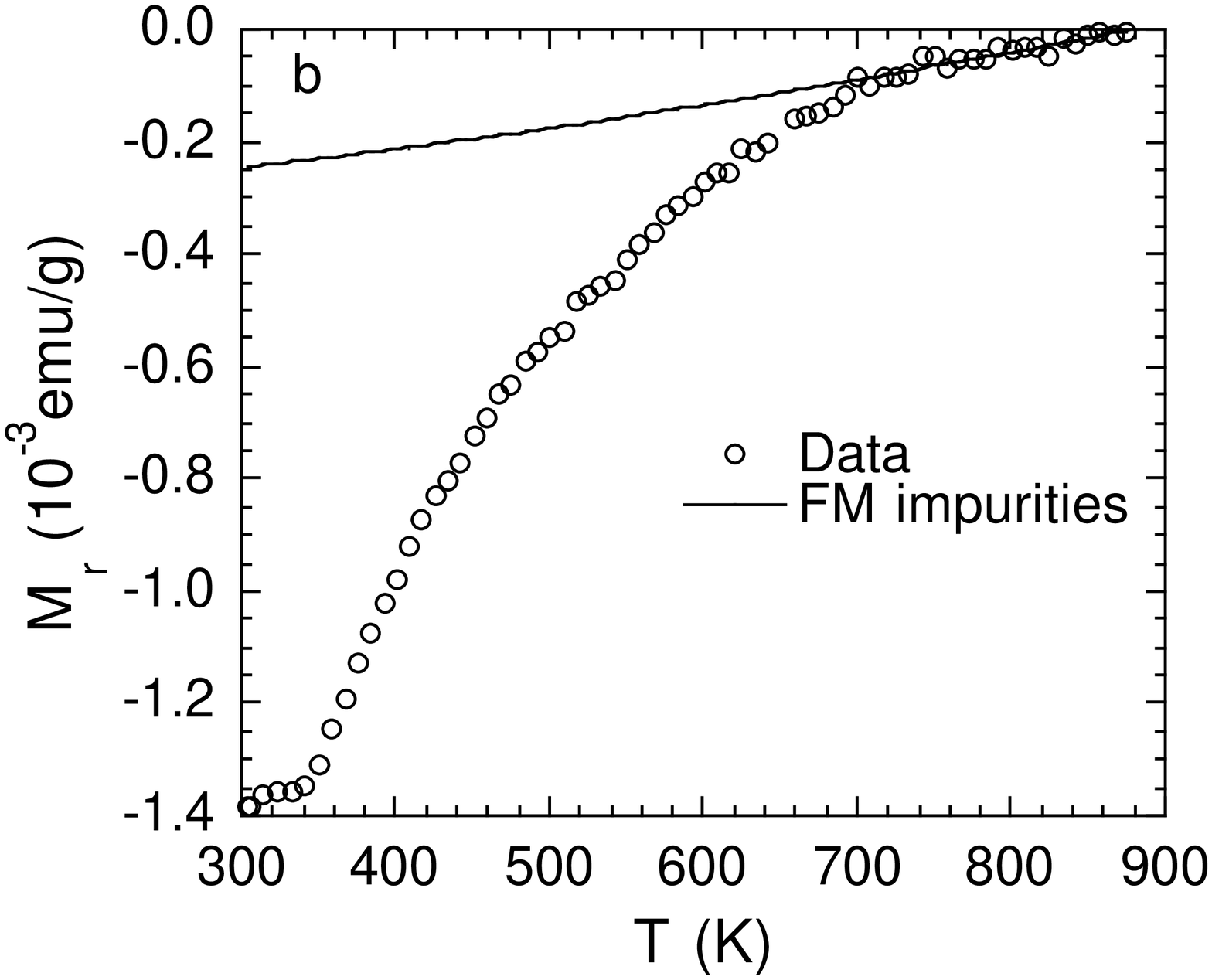}}
\vspace{0.3cm}
\caption[~]{a) The temperature dependence of the warm-up magnetization 
in a field of 0.08 Oe for sample CVD2.  b) The temperature dependence of the remanent magnetization 
for sample CVD2.  The solid line represents the ferrimagnetic remanence 
due to the Fe$_{3}$O$_{4}$ magnetic impurities.  }
\end{figure}

Fig.~17a shows the temperature dependence of the field-cooled 
susceptibility in a field of 0.08 Oe for sample CVD2.  The data 
appear 
to indicate two ferromagnetic/ferrimagnetic transitions at about 864~K 
and 1028~K, respectively, although the large data scattering may mimic a single 
transition at about 1000~K.  The former 
transition should be related to the ferrimagnetic ordering of the 
Fe$_{3}$O$_{4}$ impurities while the latter transition is associated 
with the ferromagnetic ordering of the Fe impurities.  The Fe 
impurities may be reduced from some Fe$_{3}$O$_{4}$ impurities when 
the sample is heated to 1100~K in a high vacuum.

Fig.~17b shows the temperature dependence of the ferrimagnetic 
component of the susceptibility in the field of 0.08 Oe.  This 
component is obtained by subtracting a constant paramagnetic 
term$-$evident between 1030 K and 1100 K$-$from the total 
susceptibility.  The solid line is the fitted curve by Eq.~2 with a 
fixed parameter $T_{C}$ = 1028 K and two fitting parameters: $\chi_{FM}(0)$ = 0.00148 emu/g and $p$ = 2.67.  The small $p$ value for sample 
CVD2 is consistent with the fact that this sample has two magnetic 
transitions which, due to the large data scattering, could resemble a 
single broad transition.

\begin{figure}[htb] 
 \ForceWidth{7cm}
\centerline{\BoxedEPSF{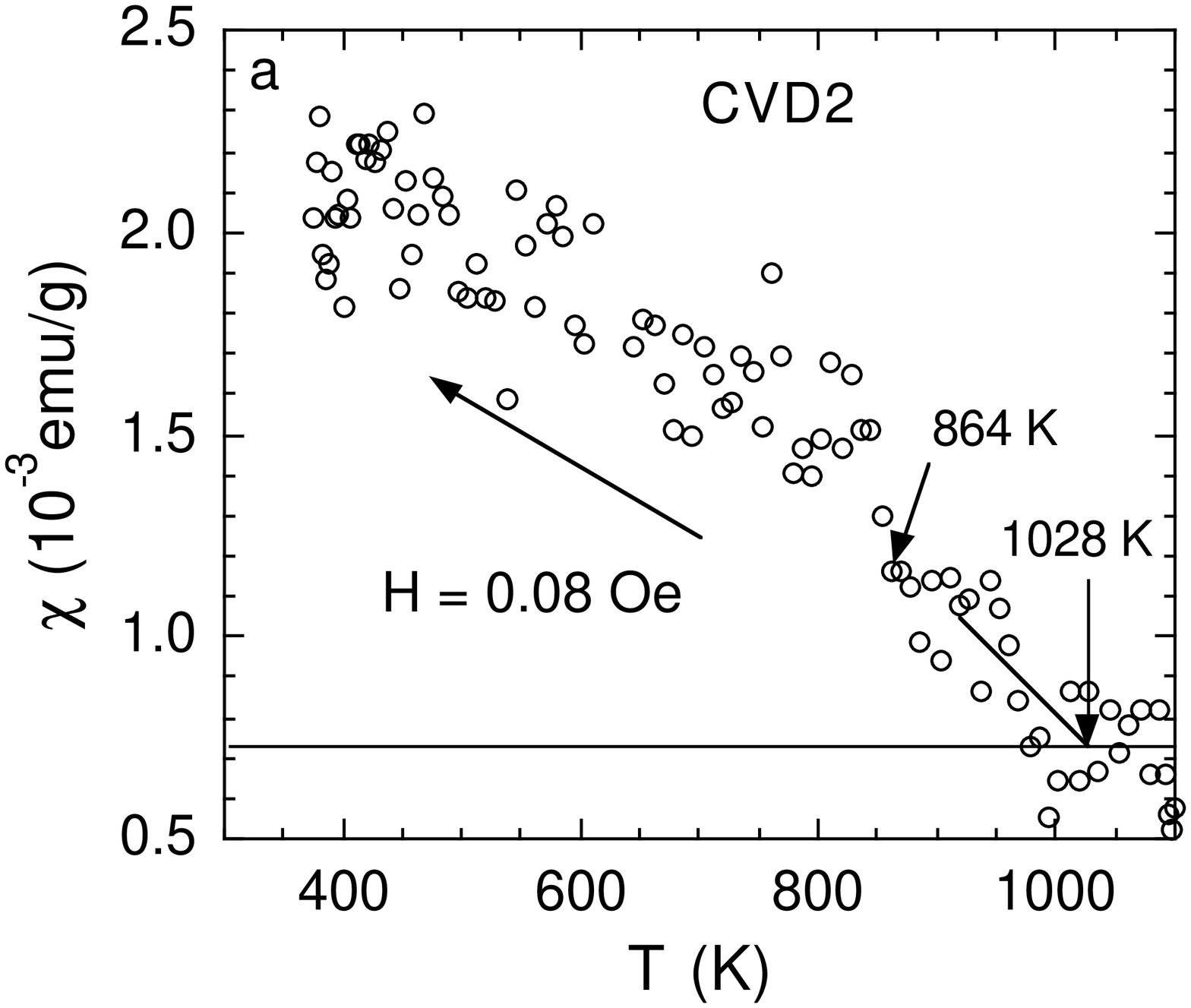}}
\vspace{-0.1cm}
\ForceWidth{7cm}
\centerline{\BoxedEPSF{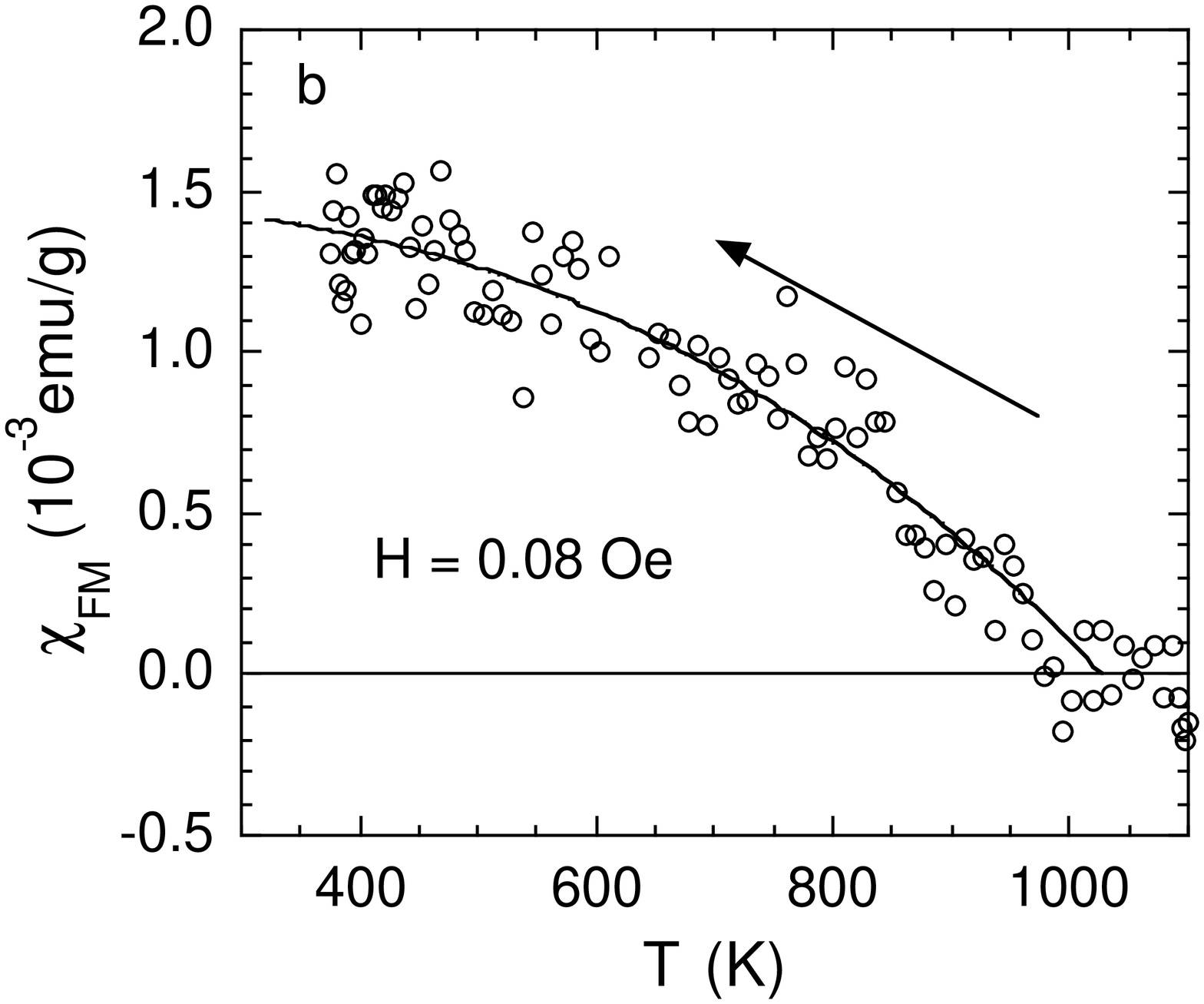}}
\vspace{0.3cm}
\caption[~]{a) The temperature dependence of the field-cooled 
susceptibility in a field of 0.08 Oe for sample CVD2.  b) The 
temperature dependence of the ferrimagnetic component of the 
susceptibility in the field of 0.08 Oe.  The solid 
line is the fitted curve by Eq.~2 with a fixed parameter $T_{C}$ = 1028 K and 
two fitting parameters: $\chi_{FM}(0)$ = 0.00148 emu/g and $p$ = 2.67.}
\end{figure}

Comparing Fig.~17b with Fig.~13c, we see that the ferrimagnetic 
susceptibility at 600 K for sample CVD2 is about a 
factor of 5 smaller than that for sample CVD1. This is consistent 
with the fact that the concentration of magnetic impurities in sample 
CVD1 is larger than that for sample CVD2 by a factor of about 5.  On 
the other hand, comparing Fig.~16b with Fig.~12b, we find that the 
ferrimagnetic remanence for sample CVD1 (from $-$200 Oe to $-$0.06 Oe) is larger than 
that for sample CVD2 (from $-$200 Oe to 0.08 Oe) by a factor of about 8. 
This discrepancy can be resolved if we consider the fact that the 
ending fields in these two cases are different. From the lower part 
of the magnetic 
hysteresis loop of Fe$_{3}$O$_{4}$, one can see that 
the magnitude of the low-field $M_{r}$ decreases linearly with the 
increase of the ending field. The 
slope of the linear decrease line is approximately equal to the 
low-field susceptibility $\chi_{FM}(0)$. This implies that when the ending field decreases 
from 0.08 Oe to $-$0.06 Oe, the magnitude of $M_{r}$ will increase by 
$\Delta H\chi_{FM} (0)$ = $0.14\chi_{FM}(0)$.
Using $\chi_{FM}(0)$ = 0.00148 
emu/g, we find that 
the magnitude of $M_{r}$ for sample CVD2 is  
(0.000307 + 0.14$\times$0.00143) emu/g = 5.1$\times$10$^{-4}$ emu/g 
if the ending field is $-$0.06 Oe.  Therefore, for the same ending 
field of $-$0.06 Oe, the ferrimagnetic 
remanence for 
sample CVD2 is smaller than that for sample CVD1 by a factor of about~5. 

\begin{figure}[htb] 
 \ForceWidth{7cm}
\centerline{\BoxedEPSF{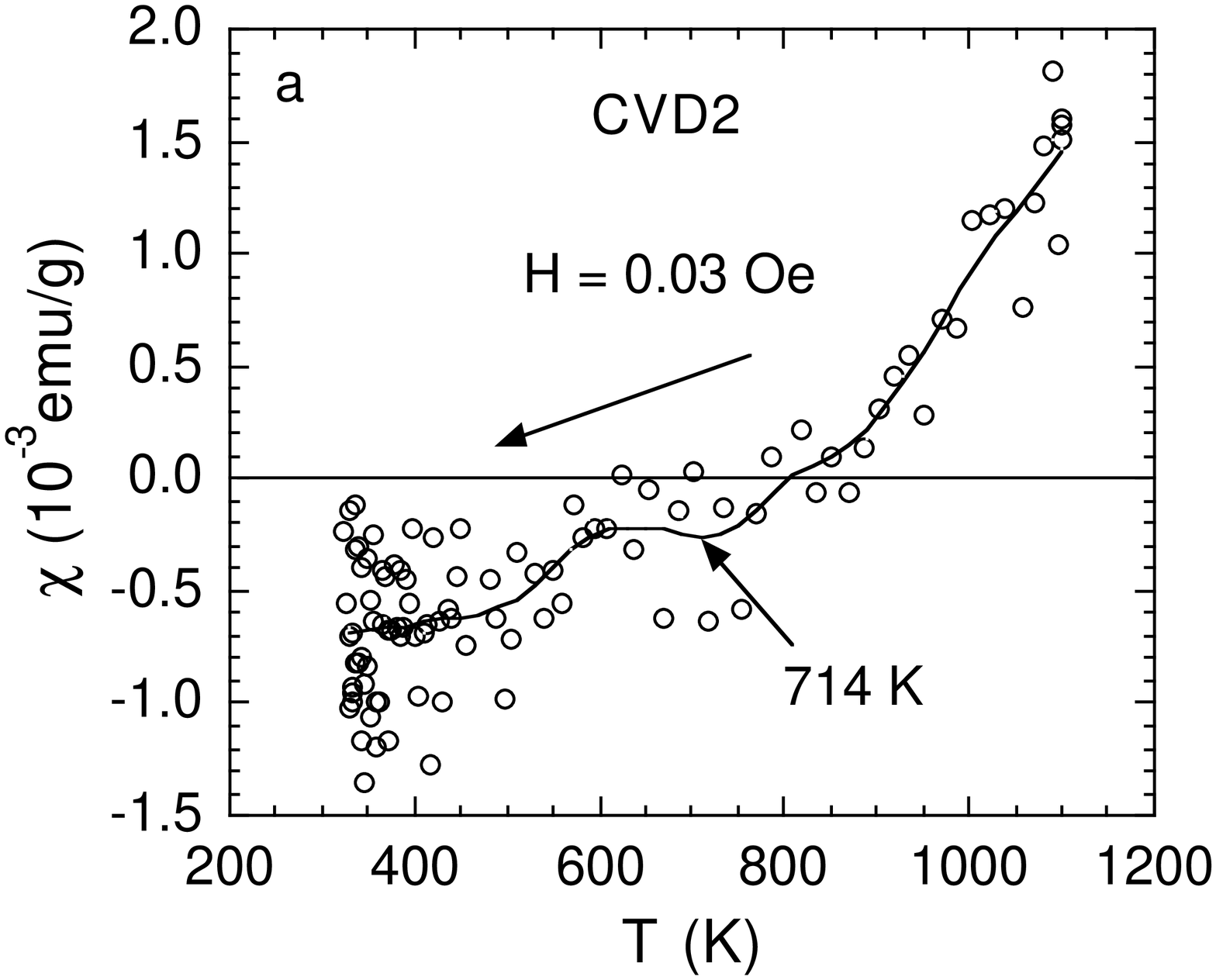}}
\vspace{-0.1cm}
 \ForceWidth{7cm}
\centerline{\BoxedEPSF{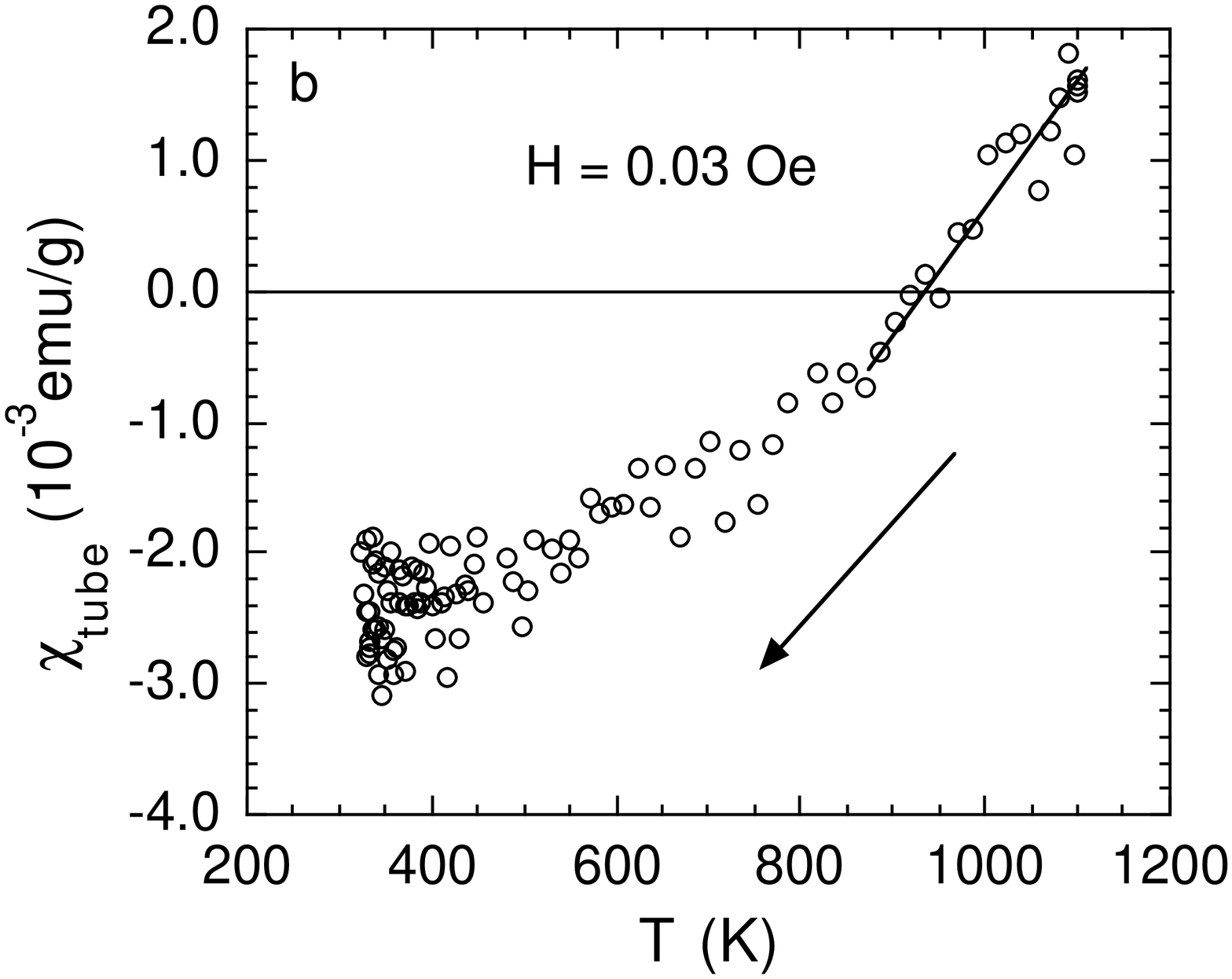}}
	\vspace{0.3cm}
\caption[~]{a) The temperature dependence of the susceptibility in a field 
of 0.03 Oe for sample CVD2.  The solid line is an averaged curve. The susceptibility at 330 K has a large negative value 
(about $-$6.9$\times$10$^{-4}$ emu/g), corresponding to about 1.9$\%$
of the full Meissner effect. b) 
The temperature dependence of the intrinsic susceptibility of the 
nanotubes.  The data are obtained by subtracting 
the ferrimagnetic/ferromagnetic contribution from the total 
susceptibility.  The intrinsic diamagnetic 
susceptibility of the nanotubes is about $-$2.5$\times$10$^{-3}$ emu/g at 
330 K, corresponding to about 7$\%$ of the
full Meissner effect.}
\end{figure} 

Fig.~18a shows the field cooled susceptibility in a field of 0.03 Oe 
for sample CVD2.  The sample was taken out of the VSM after the 
field-cooled measurement in 0.08 Oe was done.  The sample was then 
reinserted 
into the sample chamber with a field of 0.03 Oe, heated up to 1100 K 
and cooled down in the same field.  It is interesting that the 
susceptibility drops by about 2.2$\times$10$^{-3}$ emu/g when the 
temperature is lowered from 1100 K to 330 K.  This implies that the 
magnitude of the diamagnetic component at 330 K is at least 
2.2$\times$10$^{-3}$ emu/g, corresponding to about 6$\%$ of the
full Meissner effect.  More remarkably, the susceptibility at 330 K 
has a large  
negative value (about $-$6.9$\times$10$^{-4}$ emu/g), 
corresponding to about 1.9$\%$ of the
full Meissner effect.  From the averaged curve of Fig.~18a, we see a 
dip-like feature at about 700 K which corresponds to the onset of 
intergrain Josephson coupling.  Coinciding with this temperature, we 
have clearly shown a large superconducting remanence up to 700 K for 
sample CVD2 (see Fig.~16b).

We can estimate the ferrimagnetic/ferromagnetic contribution in the 
field 
of 0.03 Oe for sample CVD2 using the field dependence of the 
ferrimagnetic 
susceptibility seen in sample CVD1 (see Fig.~13c) and the temperature 
dependence of the ferrimagnetic/ferromagnetic susceptibility in 0.08 
Oe for sample CVD2.  From Fig.~13c, we have found that $\chi_{FM}$ 
$\propto$ $H^{-0.23}$.  This implies that $\chi_{FM}$ in 0.03 Oe is 
larger than that in 0.08 Oe by a factor of 1.25.  Therefore, in the 
field of 0.03 Oe, $\chi_{FM}(T) = 0.00185[1 - (T/1028)^{2.67}]$ emu/g.  
Then the intrinsic susceptibility of the carbon nanotubes is obtained 
by subtracting the ferrimagnetic/ferromagnetic contribution from the 
total susceptibility.

Fig.~18b shows the temperature dependence of 
the intrinsic susceptibility of the nanotubes in sample CVD2. We can clearly see that 
the temperature dependence of the intrinsic susceptibility for this 
CVD prepared sample is similar to that for the AD prepared sample (see Fig.~7b). 
The intrinsic 
susceptibility for this CVD prepared sample decreases by 
4.2$\times$10$^{-3}$ emu/g when the temperature is lowered from 1100 
K 
to 330 K, corresponding to about 12$\%$ of the
full Meissner effect.  It is remarkable that the magnitude of the 
intrinsic susceptibility drop for this CVD sample is almost the same as that 
(4.5$\times$10$^{-3}$ emu/g) for the AD prepared sample.  This 
consistency suggests that the observed large susceptibility drops in 
these 
two samples having very different masses and magnetic impurity 
concentrations are intrinsic to the nanotubes.  Moreover, the 
intrinsic 
diamagnetic susceptibility 
of the nanotubes is  $-$2.5$\times$10$^{-3}$ emu/g at 330 K, 
corresponding to about 7$\%$ of the
full Meissner effect. Such a large FC diamagnetic susceptibility is 
{\em only} 
consistent with superconductivity.

Next we argue that the large paramagnetic susceptibility at 1100 K 
(1.6$\times$10$^{-3}$ emu/g) observed in this CVD sample cannot arise 
from magnetic contaminants.  Since the warm-up susceptibility in 0.03 
Oe coincides with the cool-down susceptibility in the temperature region 
between 1000 and 1100 K, there should be negligible magnetic 
impurities having Curie temperatures higher than 1000 K and 
cocervities larger than 0.03 Oe.  It is highly unlikely that the 
sample contains magnetic impurities with Curie temperatures higher 
than 1100 K and with cocervities smaller than 0.03 Oe.  Furthermore, 
the susceptibility at 1100 K is inversely proportional to the field 
in the 
field region between 0.03-0.08 Oe (see Fig.~17a and Fig.~18), which 
is 
difficult to explain in terms of magnetic contaminations.  
Nevertheless, our 
conclusion about the existence of hot superconductivity in the 
nanotubes is independent of how one interprets the large paramagnetic 
susceptibility.

\section{Concluding remarks}

We report conclusive evidence for granular room-temperature 
superconductivity in multiwalled carbon nanotube mat samples with an 
intragrain transistion temperature slightly higher than 1200 K and 
intergrain transition temperatures of 533-700 K.  We show numerous 
signatures of room-temperature superconductivity: 1. a direct diamagnetic 
Meissner fraction of 1.9$\%$ at room temperature (this exceeds the orbital diamagnetic component 
by at least
two orders of
magnitude) 2. an inferred diamagnetic fraction of 14$\%$ at room 
temperature 3. coincidence of the 
resistive and magnetic transitions 4. existence of superconducting persistent
currents (superconducting remanence) up to 700 K  5.~great field 
sensitivity of the intergrain transition temperature and 6.~very 
large enhancement in the low field ferrimagnetic/ferromagnetic 
susceptibility of magnetic impurities up to 1030 K$-$consistent with 
superconductivity above 1030 K in the nanotubes.

Another important superconductivity signature is the zero resistance state. We 
have clearly shown the existence of the superconducting remanence 
up to 700 K, implying that a zero resistance state has been achieved 
below 700 K in some parts of the mat samples, where both the on-tube and 
intergrain-Josephson-junction resistances approach zero. Further, a 
negligible or zero on-tube resistance has been seen in many unprocessed 
individual MWNTs at room temperature \cite{Frank,Pablo,Heer}.  This 
very small on-tube resistivity has 
been interpreted as ``ballistic transport". Using a realistic 
electron-phonon coupling strength in carbon nanotubes, Zhao \cite{Zhaorev4} has 
shown that, for a non-superconducting SWNT with a diameter of 15 nm, the on-tube 
resistance per unit length at room 
temperature is at least 367 $\Omega$/$\mu$m, which is larger than the 
measured values \cite{Heer} by at least one order of magnitude. The 
finite but very small on-tube resistance observed in individual MWNTs 
is a natural consequence of quantum phase slips in 
quasi-one-dimensional superconductors. Bundling individual MWNTs greatly 
suppresses quantum phase slips, leading to a 
nearly zero-resistance state in some parts of the mat samples.

Finally, this work is also supported by a huge chorus of facts. The 
single-particle gap and transition temperature$-$determined 
independently from Raman data of a SWNT mat sample$-$yield a ratio 
of 1.83 (Ref.~\cite{Zhaorev2}), which is close to the BCS 
prediction (1.76) for weak coupling. The tunneling spectra also evidence 
unique superconducting signatures: The single-particle tunneling gap is found 
to be at least 125 meV in an individual MWNT with a diameter of 30 nm 
(Ref.~\cite{Zhaorev4}).  These facts along with the present magnetic 
data provide compelling evidence for room-temperature 
superconductivity in carbon nanotubes.
~\\
~\\
\noindent
{\bf Acknowledgment:} We thank Quantum Design (J. O'Brien, N.  R. 
Dilley, J.  J.  Cherry, R.  Fox and Dan Polancic) for hospitality, 
use of 
the VSMs and technical assistance.

~\\ 
 ~\\
*Correspondence should be addressed to gzhao2@calstatela.edu.

\end{document}